\documentclass[12pt,a4paper]{article}
\usepackage{amssymb,enumitem,booktabs,subfig,xcolor,todonotes,microtype,setspace,paralist,fullpage, makecell, longtable, bm}
\usepackage[top=1.4in, bottom=1.4in, left=1.35in, right=1.35in]{geometry}
\usepackage{natbib}
\usepackage{url} 
\usepackage[pdftex,colorlinks=true,hypertexnames=false]{hyperref}
\definecolor{darkblue}{rgb}{0,0,.6}
\hypersetup{citecolor=darkblue,linkcolor=darkblue,urlcolor=darkblue}
\usepackage{amsmath}
\usepackage{amsfonts}
\usepackage{epsfig}
\usepackage{graphics}
\usepackage{palatino,eulervm}
\usepackage{graphicx}
\usepackage{lscape}
\usepackage{rotating}
\usepackage[linewidth=1pt]{mdframed}
\allowdisplaybreaks[4]
\DeclareMathOperator*{\argmin}{arg\,min}
\DeclareMathOperator*{\argmax}{arg\,max}

\providecommand{\U}[1]{\protect\rule{.1in}{.1in}}

\graphicspath{{plots/}}

\usepackage{amsthm,thmtools}
\usepackage{mathrsfs}
\declaretheorem{theorem}

\makeatletter
\def\th@newremark{\th@remark\thm@headfont{\bfseries}}
\makeatletter
\theoremstyle{newremark}

\newtheorem{prop}{Proposition}

\newtheorem{assumption}{Assumption}
\declaretheoremstyle[
  spaceabove=6pt, spacebelow=6pt,
  headfont=\bfseries,
  notefont=\mdseries, notebraces={(}{)},
bodyfont=\normalfont,
  postheadspace=0.5em,
]{mystyle}

\begin{document}

\title{Detection and Estimation of Structural Breaks in High-Dimensional Functional Time Series}
\author{{\normalsize Degui Li\thanks{Department of Mathematics, University of York, UK. },\ \ \ Runze Li\thanks{Department of Statistics, Penn State University, US. The corresponding author, rzli@psu.edu},\ \ \ Han Lin Shang\thanks{Department of Actuarial Studies and Business Analytics, Macquarie University, Australia}}\\
{\normalsize\em University of York, Penn State University and Macquarie University}}

\date{\normalsize This version: \today}

\maketitle

\centerline{\bf Abstract}

\medskip

In this paper, we consider detecting and estimating breaks in heterogeneous mean functions of high-dimensional functional time series which are allowed to be cross-sectionally correlated and temporally dependent. A new test statistic combining the functional CUSUM statistic and power enhancement component is proposed with asymptotic null distribution theory comparable to the conventional CUSUM theory derived for a single functional time series. In particular, the extra power enhancement component enlarges the region where the proposed test has power, and results in stable power performance when breaks are sparse in the alternative hypothesis. Furthermore, we impose a latent group structure on the subjects with heterogeneous break points and introduce an easy-to-implement clustering algorithm with an information criterion to consistently estimate the unknown group number and membership. The estimated group structure can subsequently improve the convergence property of the post-clustering break point estimate. Monte-Carlo simulation studies and empirical applications show that the proposed estimation and testing techniques have satisfactory performance in finite samples.

\medskip

\noindent{\em Keywords}: clustering, CUSUM, functional time series, power enhancement, structural breaks.

\newpage


\section{Introduction}\label{sec1}
\renewcommand{\theequation}{1.\arabic{equation}}
\setcounter{equation}{0}

Modelling functional time series, time series of random functions defined within a finite interval, has became one of the main frontiers of developments in time series models. Various functional linear and nonlinear time series models have been proposed and extensively studied in the past two decades \citep[e.g.,][]{B00, HK10, HK12, HHR13, LRS20}. These models together with relevant methodologies have been applied to various fields such as biology, demography, economics, environmental science and finance. However, the model frameworks and methodologies developed in the aforementioned literature heavily rely on the stationarity assumption, which is often rejected when testing the functional time series data in practice. For example, \cite{HKR14} find evidence of nonstationarity for intraday price curves of some stocks collected in the US market; \cite{ARS18} reject the null hypothesis of stationarity for the temperature curves collected in Australia; and \cite{LRS23} reveal evidence of nonstationary feature for the functional time series constructed from the age- and sex-specific life-table death counts. It thus becomes imperative to test whether the collected functional time series are stationary.

\smallskip

The primary interest of this paper is to test whether there exist structural breaks in the mean function over time and subsequently estimate locations of breaks if they do exist. There have been increasing interests on detecting and estimating structural breaks in functional time series. Broadly speaking, there are two types of detection techniques. One is to first reduce the infinite dimension of functional data to a finite dimension via the classic functional principal component analysis and then use the detection method developed for multivariate data to identify breaks \citep[e.g.,][]{AGHP09,BGHP09, ZSHW11, AK12}, and the other is a fully functional detection method without preliminary dimension reduction \citep[e.g.,][]{HKR14, STW16, ARS18}. The latter avoids possible information loss caused by the dimension reduction, thus may have more reliable numerical performance. The existing research works often focus on break detection in a single functional time series sequence. An extension to multiple functional time series is straightforward, but a further extension to high-dimensional functional time series is challenging. In fact, so far as we know, there is virtually no work on break detection and estimation in high-dimensional functional time series with dimension comparable to the sample size.  

\smallskip

In this paper, we consider structural break detection within a flexible high-dimensional functional time series model framework, allowing the functional time series to be cross-sectionally correlated over subjects and temporally dependent over time. A novel test statistic combining the CUSUM statistic with the power enhancement (PE) component is proposed to detect breaks in heterogeneous mean functions. The asymptotic null distribution of the proposed test statistic is comparable to that of the conventional CUSUM statistic developed for a single functional time series \citep[e.g.,][]{HKR14, ARS18}. In practice, it is often the case that breaks are sparse in the alternative hypothesis, i.e., breaks only occur in a small number of functional time series, or that the break locations may vary over subjects. Consequently, the traditional CUSUM test statistic would have low powers. To address this problem, an extra PE component is added to the test statistic. The PE technique is introduced by \cite{FLY15} to improve the power performance of high-dimensional tests for sparse alternatives and is further studied by \cite{KP19}. We will show that incorporating the PE component in the break test enlarges the region where the proposed test has power, while avoids size distortion. 

\smallskip

When a large number of functional time series are available, under the alternative hypothesis (with functional structural breaks), it is natural to split all the subjects into two groups: subjects of functional time series without breaks, and those with breaks. For the latter, we further estimate the unknown break points and derive a uniform approximation rate which is nearly optimal up to a logarithmic factor. In addition, we assume that there exists a latent group structure on the heterogeneous break points, i.e., subjects in each group have a common break point, whereas the break points are distinct between groups, and develop an easy-to-implement clustering method with an information criterion to consistently estimate the latent structure including the group membership and number of distinct break points. This complements the recent literature on cluster analysis of functional data which are often assumed to be independent and identically distributed (i.i.d.) or stationary and weakly dependent \citep[e.g.,][]{TK03, FV06, CL07, PM08, DH12, DHP19}. Furthermore, we estimate the homogenous break point within each group and show that the consistency property can be improved for the estimated break point by making use of the estimated group structure. 

\smallskip

The Monte-Carlo simulation studies show that the proposed test with the extra PE component achieves the power enhancement when structural breaks are relatively sparse while avoids size distortion in finite samples. We also compare the numerical performance between different choices of high-criticism thresholding parameters. With the latent group structure on change points, the proposed clustering algorithm and information criterion can accurately identify the cluster number and membership. Furthermore, the pooled CUSUM method using the estimated cluster structure can substantially improve the estimation accuracy of the common break points. The developed break detection and estimation methods are applied to Dow Jones Industrial Average constituent stocks and the age-specific mortality rates collected in $32$ countries to test the existence of structural breaks and estimate the break locations. The empirical analysis reveals that there exists one common break for either of the two real data sets. 

\smallskip

The rest of the paper is organised as follows. Section~\ref{sec2} presents the model framework, hypothesis testing problem and some fundamental assumptions. Section~\ref{sec3} introduces the test statistic together with its asymptotic properties under both the null and alternative hypotheses. Section~\ref{sec4} considers the break point estimation and identifies the latent structure on the heterogeneous break points. Sections~\ref{sec5} and~\ref{sec6} report the simulation and empirical studies, respectively. Section~\ref{sec7} concludes the paper. Proofs of the main asymptotic theorems are given in Appendix A whereas proofs of some technical lemmas and propositions are available in a supplement. Throughout the paper, we let ${\mathscr H}$ be the Hilbert space defined as a set of measurable functions $f(\cdot)$ on a bounded set ${\mathbb C}$ such that $\int_{\mathbb C} f^2(u)du<\infty$. The inner product in ${\mathscr H}$ is defined as $\langle f_1,f_2\rangle=\int_{\mathbb C}f_1(u)f_2(u)du$. For $f\in{\mathscr H}$, we define the $L_2$-norm $\Vert f\Vert:=\Vert f\Vert_2=\langle f,f\rangle^{1/2}=\left[\int_{\mathbb C}|f(u)|^2du\right]^{1/2}$ and more generally the $L_p$-norm $\Vert f\Vert_p=\left[\int_{\mathbb C}|f(u)|^pdu\right]^{1/p}$, $p\geq1$. Let ${\mathscr L}({\mathscr H})$ be the space of continuous linear operators from ${\mathscr H}$ to ${\mathscr H}$ equipped with the operation norm defined by $\Vert{\mathcal L}\Vert_O=\sup_{f\in{\mathscr H}}\left\{\Vert{\mathcal L}(f)\Vert: \Vert f\Vert\leq1\right\}$. For each ${\mathcal L}\in{\mathscr L}({\mathscr H})$, its adjoint ${\mathcal L}^\prime$ is defined via $\langle {\mathcal L}f_1,f_2\rangle=\langle f_1,{\mathcal L}^\prime f_2\rangle$ for any $f_1,f_2\in{\mathscr H}$. Let $\stackrel{d}\longrightarrow$, $\stackrel{P}\longrightarrow$ and $\Rightarrow$ denote convergence in distribution, convergence in probability and weak convergence, respectively.


\section{Model and assumptions}\label{sec2}
\renewcommand{\theequation}{2.\arabic{equation}}
\setcounter{equation}{0}

Suppose that we collect a sequence of functional observations ${\mathbf X}_t=(X_{1t},\cdots,X_{Nt})^{^\intercal}$, $t=1,\cdots,T$, where $X_{it}=\left(X_{it}(u): u\in{\mathbb C}\right)$. For the $i$-th subject, $X_{it}$, $t=1,\cdots,T$, are generated from the following model with a possible break in the mean function:
\begin{equation}\label{eq2.1}
X_{it}=\mu_i+\delta_i I\left(t>\tau_i\right)+\epsilon_{it},
\end{equation}
where $\mu_i=(\mu_i(u): u\in{\mathbb C})$ is the pre-break mean function, $\delta_i=(\delta_i(u): u\in{\mathbb C})$ is the jump function, $\tau_i$ is the break point, $I(\cdot)$ is the indicator function and $\epsilon_{it}=(\epsilon_{it}(u): u\in{\mathbb C})$ is stationary over the temporal dimension. The unknown functional components $\mu_i$ and $\delta_i$ as well as the break points $\tau_i$ are allowed to vary over $i$, reflecting heterogeneity of functional time series over subjects. Our primary interest lies in the following hypothesis testing problem:
\begin{equation}\label{eq2.2}
H_0:\ \delta_i=0,\ i=1,\cdots,N,\ \ \ \ {\rm versus}\ \ \ \ H_A:\ \delta_i\neq 0\ {\rm for\ some}\ i.
\end{equation}
Under the null hypothesis $H_0$, model~\eqref{eq2.1} reduces to $X_{it}=\mu_i+\epsilon_{it}$, which is stationary over $t$. We aim to test whether to reject $H_0$ and estimate the break points $\tau_i$ if $H_0$ is rejected. The functional errors $\epsilon_{it}$ satisfy the following regularity conditions.

\renewcommand{\theassumption}{\arabic{assumption}}
\setcounter{assumption}{0}

\begin{assumption}\label{ass:1}

{\em (i)\ Let 
\begin{equation}\label{eq2.3}
\epsilon_{it}=\sum_{j=0}^\infty{\mathbf A}_{ij}\eta_{i,t-j},
\end{equation}
where $\eta_{it}=(\eta_{it}(u): u\in {\mathbb C})$ are i.i.d. random elements in ${\mathscr H}$ with mean zero and positive definite covariance function $\Omega_{i}(u,v)$, $u,v\in{\mathbb C}$, and ${\mathbf A}_{ij}$ are continuous linear operators with the operator norm satisfying 
\begin{equation}\label{eq2.4}
\sum_{j=0}^\infty j  \left(\max_{1\leq i\leq N} \Vert {\mathbf A}_{ij}\Vert_O\right)<C_A,
\end{equation}
where $C_A$ is a positive constant which does not depend on $N$.}

{\em (ii) Let 
\[\widetilde{\eta}_{Nt}=\frac{1}{\sqrt{N}}\sum_{i=1}^N {\mathbf A}_i \eta_{is},\ \ {\mathbf A}_i=\sum_{j=0}^\infty{\mathbf A}_{ij}.\]
There exists a positive definite integral operator $\widetilde{\boldsymbol\Omega}$ with the kernel satisfying
\begin{equation}\label{eq2.5}
\widetilde{\Omega}(u,v)=\lim_{N\rightarrow\infty} {\sf E}\left[\widetilde{\eta}_{Nt}(u) \widetilde{\eta}_{Nt}(v)\right].
\end{equation}
Furthermore, for any $i$, ${\sf E}\left[\exp\left\{c_\eta\Vert\eta_{it}\Vert^2\right\}\right]<\infty$ with $c_\eta$ being a positive and bounded constant, and for any sequence of continuous linear operators ${\mathbf B}_i$,}
\begin{equation}\label{eq2.6}
{\sf E}\left[\left\Vert\frac{1}{\sqrt{N}}\sum_{i=1}^N{\mathbf B}_i\eta_{it}\right\Vert_O^{2+\iota} \right]=O\left(\max_{1\leq i\leq N}\Vert{\mathbf B}_i\Vert_O^{2+\iota}\right),\ \ 0\leq \iota\leq 1.
\end{equation}

\end{assumption}

From Assumption~\ref{ass:1}(i), the functional linear process $\epsilon_{it}$ defined in~\eqref{eq2.3} with coefficient operators satisfying~\eqref{eq2.4} is stationary and short-range dependent over time $t$, but its distribution is allowed to vary over $i$. This assumption is a natural extension of the assumptions in \cite{Ba10} and \cite{HH12} for panel time series setting. Assumption~\ref{ass:1}(ii) shows that the functional time series may be weakly cross-sectionally correlated over $i$. The moment restriction in~\eqref{eq2.6} is a high-level condition which is easy to verify when $\eta_{it}$ are independent over $i$. In the latter case, we may further show that 
\[\widetilde{\boldsymbol\Omega}=\lim_{N\rightarrow\infty}\frac{1}{N}\sum_{i=1}^N{\mathbf A}_i{\boldsymbol\Omega}_i{\mathbf A}_i^\prime,\] where ${\boldsymbol\Omega}_i$ is the integral operator with $\Omega_i(u,v)$ as the kernel. As $\widetilde{\boldsymbol\Omega}$ is positive definite, we may conduct an eigenanalysis and find pairs of non-negative eigenvalues and eigenvectors $(\lambda_{k},\psi_{k})$ (with eigenvalues arranged in an non-increasing order), $k=1,2,\cdots$, such that
\begin{equation}\label{eq2.7}
\widetilde{\boldsymbol\Omega}(\psi_{k})(u)=\lambda_{k}\psi_{k}(u).
\end{equation}
The exponential moment condition ${\sf E}\left[\exp\left\{c_\eta\Vert\eta_{it}\Vert^2\right\}\right]<\infty$ in Assumption~\ref{ass:1}(ii) facilitates the truncation technique and application of the concentration inequality for random elements in ${\mathscr H}$ \citep[e.g.,][]{B00, BLM16} in proofs of the main asymptotic results. 


\section{Testing structural breaks}\label{sec3}
\renewcommand{\theequation}{3.\arabic{equation}}
\setcounter{equation}{0}

Letting $\nu_{it}=\mu_i+\delta_i I\left(t>\tau_i\right)$, we may re-write model~\eqref{eq2.1} as 
\begin{equation}\label{eq3.1}
X_{it}=\nu_{it}+\epsilon_{it},\ \ t=1,\cdots,T,\ \ i=1,\cdots,N.
\end{equation}
Define 
$$\widetilde{X}_t=\frac{1}{N}\sum_{i=1}^N X_{it},\ \ \widetilde{\nu}_t=\frac{1}{N}\sum_{i=1}^N\nu_i\ \ {\rm and}\ \ \widetilde{\epsilon}_t=\frac{1}{N}\sum_{i=1}^N \epsilon_{it},$$
where we suppress their dependence on $N$. From the model formulation~\eqref{eq3.1}, we have
\begin{equation}\label{eq3.2}
\widetilde{X}_t=\widetilde{\nu}_t+\widetilde{\epsilon}_t,\ \ t=1,\cdots,T.
\end{equation}
Note that, under $H_0$, the time-varying mean function $\widetilde{\nu}_t$ becomes $\widetilde{\mu}=\frac{1}{N}\sum_{i=1}^N\mu_{i}$ which is time-invariant and model~\eqref{eq3.2} thus reduces to
\[\widetilde{X}_t=\widetilde{\mu}+\widetilde{\epsilon}_t,\ \ t=1,\cdots,T.\]
In order to test $H_0$ in~\eqref{eq2.2}, a naive idea is to construct the CUSUM test statistic using $\widetilde{X}_{t}$. Define the functional CUSUM statistic as 
\begin{equation}\label{eq3.3}
\widetilde{Z}_{NT}(x;u)=\sqrt{N/T}\left(\widetilde{S}_{NT}(x;u)-\frac{\lfloor Tx\rfloor}{T}\widetilde{S}_{NT}(1;u)\right)\ \ {\rm with}\ \  \widetilde{S}_{NT}(x;u)=\sum_{s=1}^{\lfloor Tx\rfloor}\widetilde{X}_{s}(u),
\end{equation}
where $0\leq x\leq 1$, $u\in{\mathbb C}$ and $\lfloor \cdot\rfloor$ denotes the floor function, and subsequently construct the test statistic via 
\begin{equation}\label{eq3.4}
Z_{NT}=\sup_{0\leq x\leq 1}\int_{\mathbb C}\widetilde{Z}_{NT}^2(x;u)du=\max_{1\leq t\leq T}\int_{\mathbb C}\widetilde{Z}_{NT}^2(t/T;u)du.
\end{equation}
The $\sqrt{N}$-rate in the definition of $\widetilde{Z}_{NT}(x;u)$ is appropriate when $X_{it}$ and $\epsilon_{it}$ are weakly dependent (or independent as a special case) over $i$. If the functional time series are strongly dependent over $i$, a different normalisation rate may be required. Under some regularity conditions, we may show that $Z_{NT}$ has the same asymptotic null distribution as the classic CUSUM test statistic for a single functional time series, see Theorem~\ref{thm:1} below. However, the test statistic based on the model formulation~\eqref{eq3.2} suffers from the low-power issue as $\widetilde{\nu}_t$ may be time-invariant even when there are significant breaks in the subject-specific mean functions $\nu_{it}$ for some indices $i$. For example, in the case of common break ($\tau_i\equiv \tau$), when $N$ is even, $\delta_i=\delta$ for $i=1,\cdots,N/2$ and  $\delta_i=-\delta$ for $i=N/2+1,\cdots,N$, we have $\widetilde\nu_t=\widetilde\mu$. Consequently, $\widetilde{X}_{t}=\tilde{\mu}+\widetilde{\epsilon}_{t}$ under both $H_0$ and $H_A$ and the CUSUM test statistic in~\eqref{eq3.4} would have no power. If $\delta_i\neq 0$ only holds for a fixed number of indices $i$ (i.e., breaks are sparse), we may show that $\widetilde\nu_t\approx\widetilde\mu$, which would also result in low power of the CUSUM test.

\smallskip

To fix the low-power issue of the functional CUSUM test statistic defined in~\eqref{eq3.3} and~\eqref{eq3.4}, we next consider incorporating a PE component in construction of the break test statistic. \cite{FLY15} introduce the PE technique in high-dimensional cross-sectional tests, improving the power performance in testing sparse-type alternatives. The PE component is expected to satisfy the following three properties: (i) non-negativity, (ii) no size-distortion, and (iii) power enhancement. We aim to augment the CUSUM test statistic by adding a PE component which satisfies these three properties. Define the subject-specific functional CUSUM statistic:
\begin{equation}\label{eq3.5}
Z_{iT}(x;u)=\frac{1}{\sqrt{T}}\left[S_{iT}(x;u)-\frac{\lfloor Tx\rfloor}{T}S_{iT}(1;u)\right]\ \ {\rm with}\ \ S_{iT}(x;u)=\sum_{s=1}^{\lfloor Tx\rfloor} X_{is}(u),
\end{equation}
and the PE component:
\begin{equation}\label{eq3.6}
Z_{NT}^\diamond=\sqrt{N\vee T}\sum_{i=1}^NI\left(\sup_{0\leq x\leq 1}\int_{\mathbb C} Z_{iT}^2(x;u)du>\xi_{NT}\right),
\end{equation}
where $\xi_{NT}$ denotes a high-criticism threshold and $a\vee b=\max\{a,b\}$. The PE component $Z_{NT}^\diamond$ is clearly non-negative. We need to show that ${\sf P}\left(Z_{NT}^\diamond=0 | H_0\right)\rightarrow1$ and $Z_{NT}^\diamond$ diverges to infinity with probability approaching one ({\em w.p.a.1}) under some specific regions of the alternative hypothesis on which the CUSUM test defined in~\eqref{eq3.3} and~\eqref{eq3.4} has low power, verifying properties (ii) and (iii) of the PE component (see the proof of Theorem~\ref{thm:1}). In particular, we adopt the diverging factor $\sqrt{N\vee T}$ in the PE component rather than $\sqrt{N}$ used by \cite{FLY15}, to ensure that property (iii) can be achieved no matter $N$ is fixed or diverging. The formulation of $Z_{NT}^\diamond$ in~\eqref{eq3.6} is analogous to the sparsified CUSUM statistic proposed by \cite{CF15} for high-dimensional real-valued time series. Throughout this paper we set the threshold as $\xi_{NT}=c_\xi\ln(N\vee T)\ln\ln (N\vee T)$ with $c_\xi$ being a user-specified positive constant. In Section~\ref{sec5} below, we also consider $\xi_{NT}=c_\xi\ln(NT)\ln\ln (NT)$ in the simulation study and compare the finite-sample numerical performance between these two choices of high-criticism thresholding parameters.

\smallskip

Combining the CUSUM test statistic $Z_{NT}$ defined in~\eqref{eq3.4} and the PE component $Z_{NT}^\diamond$ defined in~\eqref{eq3.6}, we propose the following power enhanced CUSUM (PE-CUSUM) test statistic:
\begin{equation}\label{eq3.7}
\widehat{Z}_{NT}=Z_{NT}+Z_{NT}^\diamond.
\end{equation}
Let $H_A^\diamond$ be the alternative $H_A$ defined in~\eqref{eq2.2} with $\delta_i$ satisfying 
\begin{equation}\label{eq3.8}
\max_{1\leq i\leq N}\frac{T\omega_{Ti}^2\Vert\delta_i\Vert^2}{\xi_{NT}}\rightarrow\infty,\ \ \omega_{Ti}=(\tau_i/T)\wedge(1-\tau_i/T),
\end{equation}
and $\widetilde{H}_A^\diamond$ the alternative $H_A$ with $\delta_i$ satisfying~\eqref{eq3.8} or $\widetilde{\nu}_t$ defined in~\eqref{eq3.2} satisfying
\begin{equation}\label{eq3.9}
\sup_{0\leq x\leq 1}\int_{\mathbb C}\nu_{NT}^2(x,u)du\rightarrow\infty\ \ {\rm with}\ \ \nu_{NT}(x,u)=\frac{1}{\sqrt{T}}\left[\sum_{s=1}^{\lfloor Tx\rfloor}\widetilde\nu_s(u)-x\sum_{s=1}^{T}\widetilde\nu_s(u)\right].
\end{equation}
The following theorem gives the asymptotic properties of the PE-CUSUM test statistic $\widehat{Z}_{NT}$.

\renewcommand{\thetheorem}{\arabic{theorem}}
\setcounter{theorem}{0}

\begin{theorem}\label{thm:1} 

Suppose that Assumption~\ref{ass:1} is satisfied and there exists $\kappa\geq0$ such that $N=O(T^{\kappa})$. 

(i) Under $H_0$, as $N,T\rightarrow\infty$ jointly,
\begin{equation}\label{eq3.10}
\widehat{Z}_{NT}\stackrel{d}\longrightarrow\sup_{0\leq x\leq 1}\sum_{i=1}^\infty\lambda_iB_i^2(x),
\end{equation}
where $\lambda_i$, $i=1,2,\cdots$, are the eigenvalues defined in~\eqref{eq2.7} and $B_i(\cdot)$, $i=1,2,\cdots$, are independent standard Brownian bridges defined on $[0,1]$.

(ii) Under $H_A^\diamond$, as $T\rightarrow\infty$, 
\begin{equation}\label{eq3.11}
{\sf P}\left(Z_{NT}^\diamond\geq \sqrt{N\vee T}\right)\rightarrow1;
\end{equation} 
and, under $\widetilde{H}_A^\diamond$, as $N,T\rightarrow\infty$ jointly, 
\begin{equation}\label{eq3.12}
{\sf P}\left(\widehat{Z}_{NT}\geq z_\alpha\right)\rightarrow1,
\end{equation} 
where $z_\alpha$ is the upper $\alpha$-quantile of $\sup_{0\leq x\leq 1}\sum_{i=1}^\infty\lambda_iB_i^2(x)$.

\end{theorem}

\medskip

The condition $N=O(T^{\kappa})$ indicates that the number of subjects $N$ can be much larger than the time series length $T$ ($\kappa>1$). The asymptotic null distribution of the PE-CUSUM test statistic is similar to that in Theorem 2.1 of \cite{HKR14}, Corollary 1 of \cite{STW16} and Theorem 1 of \cite{ARS18}. The involvement of the PE component does not lead to size distortion, i.e., ${\sf P}\left(Z_{NT}^\diamond=0 | H_0\right)\rightarrow1$ and property (ii) of the PE component is thus satisfied. The limit distribution in~\eqref{eq3.10} relies on the unknown eigenvalues $\lambda_i$ and the standard Brownian bridges $B_i(x)$. \cite{STW16} introduce a block bootstrap method to choose the critical value of the test, whereas \cite{ARS18} suggest a Monte-Carlo simulation method. 

Theorem~\ref{thm:1}(ii) shows that the developed test is consistent under the alternative hypothesis when either~\eqref{eq3.8} or~\eqref{eq3.9} is satisfied. Without the PE component $Z_{NT}^\diamond$, the CUSUM test statistic is only consistent under the restrictive high-level condition~\eqref{eq3.9}, which can be seen as a diverging time-varying measure of the mean functions. Note that the latter condition is often violated when breaks from different subjects are cancelled out as discussed earlier in the section. By incorporating $Z_{NT}^\diamond$, the PE property is achieved, i.e., the region where the proposed test has power is enlarged from $H_A$ with~\eqref{eq3.9} to $H_A$ with either~\eqref{eq3.8} or~\eqref{eq3.9}. In particular, the condition~\eqref{eq3.8} covers the scenario of sparse breaks. It follows from~\eqref{eq3.11} that the PE property is achieved when $N$ is either fixed or diverging.


\section{Estimating break points and the latent structure}\label{sec4}
\renewcommand{\theequation}{4.\arabic{equation}}
\setcounter{equation}{0}

When $H_0$ is rejected, it is often the case that structural breaks only occur in some of the functional time series processes. Hence, we may split the index set $\{1,2,\cdots,N\}$ into 
\[{\mathcal C}_\bullet=\{ 1\leq i\leq N: \delta_i\neq 0\}\ \ {\rm and}\ \ {\mathcal C}_\circ=\{ 1\leq i\leq N: \delta_i=0\},\]
the index set with structural breaks in mean functions and that without breaks. It is natural to  estimate ${\mathcal C}_\bullet$ and ${\mathcal C}_\circ$ by 
\[
\widehat{\mathcal C}_\bullet=\left\{ 1\leq i\leq N: \sup_{0\leq x\leq 1}\int_{\mathbb C} Z_{iT}^2(x;u)du\geq\xi_{NT}\right\}\]
and
\[\widehat{\mathcal C}_\circ=\left\{ 1\leq i\leq N: \sup_{0\leq x\leq 1}\int_{\mathbb C} Z_{iT}^2(x;u)du<\xi_{NT}\right\},
\]
respectively, where $Z_{iT}(x;u)$ and $\xi_{NT}$ are defined in Section~\ref{sec3}. Subsequently, we estimate the (heterogeneous) break points $\tau_i$ by
\begin{equation}\label{eq4.1}
\widehat{\tau}_i=\argmax_{1\leq t\leq T}\int_{\mathbb C} Z_{iT}^2(t/T;u)du,\ \ i\in\widehat{\mathcal C}_\bullet.
\end{equation}
The following theorem shows that $\widehat{\mathcal C}_\bullet$ and $\widehat{\mathcal C}_\circ$ are consistent estimates and provides a uniform approximation rate for $\widehat{\tau}_i$ over $i\in{\mathcal C}_\bullet$.

\begin{theorem}\label{thm:2}

Suppose that Assumption~\ref{ass:1} is satisfied,
\begin{equation}\label{eq4.2}
\min_{i\in{\mathcal C}_\bullet}\frac{T\omega_{Ti}^2\Vert\delta_i\Vert^2}{\xi_{NT}}\rightarrow\infty
\end{equation}
with $\omega_{Ti}$ defined in~\eqref{eq3.8}, and there exists $\kappa\geq0$ such that $N=O(T^{\kappa})$. Then we have
\begin{equation}\label{eq4.3}
{\sf P}\left( \widehat{\mathcal C}_\circ={\mathcal C}_\circ,\ \widehat{\mathcal C}_\bullet={\mathcal C}_\bullet\right)\rightarrow1.
\end{equation}
If, in addition,~\eqref{eq4.2} is strengthened to
\begin{equation}\label{eq4.4}
\min_{i\in{\mathcal C}_\bullet}\left(\omega_{Ti}\Vert\delta_i\Vert\right)\geq c_\delta,
\end{equation}
where $c_\delta$ is a positive constant, we have
\begin{equation}\label{eq4.5}
\max_{i\in{\mathcal C}_\bullet}\left\vert \widehat{\tau}_i-\tau_i\right\vert=o_P\left(\left[\ln(N\vee T)\right]^{1+\zeta}\right),
\end{equation}
where $\zeta$ is an arbitrarily small positive number.

\end{theorem}

As $N=O(T^\kappa)$, the uniform approximation rate in~\eqref{eq4.5} can be simplified to $(\ln T)^{1+\zeta}$. The uniform approximation rate for the estimated break point is nearly optimal up to a logarithmic factor, and is slightly slower than the $O_P(1)$ rate derived for the single functional time series setting. We next show that the approximation rate in~\eqref{eq4.5} can be further improved by imposing a latent group structure on the break points $\tau_i$, $i\in{\mathcal C}_\bullet$.

\smallskip

Assume that there exists a partition of ${\mathcal C}_\bullet$, denoted by ${\mathcal C}(b_1),\cdots,{\mathcal C}(b_{K_0})$, such that
\begin{equation}\label{eq4.6}
\tau_i=b_k\ \ \forall\ i\in {\mathcal C}(b_k),\ \ {\mathcal C}(b_{k_1})\cap{\mathcal C}(b_{k_2})=\emptyset\ \ {\rm for}\ \ k_1\neq k_2,\ \ \cup_{k=1}^{K_0}{\mathcal C}(b_k)={\mathcal C}_\bullet,
\end{equation}
where $b_1<b_2<\cdots<b_{K_0}$ are $K_0$ distinct break points. Neither the group membership nor the number $K_0$ is known. Note that the common break setting is nested in the latent structure~\eqref{eq4.6} with $b_1=\cdots=b_{K_0}=\tau_0$ and ${\mathcal C}(\tau_0)={\mathcal C}_\bullet$. The latent structure~\eqref{eq4.6} shows that $X_{it}$ have the common break point $b_k$ when $i\in{\mathcal C}(b_k)$. 

\smallskip

With the estimated break points $\widehat{\tau}_i$ defined in~\eqref{eq4.1}, $i\in\widehat{\mathcal C}_\bullet$, we first sort them from minimum to maximum and denote the ordered points as $\widehat{\tau}^{(1)},\cdots,\widehat\tau^{(n)}$, $n=\big|\widehat{\mathcal C}_\bullet \big|$, where $|{\cal A}|$ denotes the cardinality of set ${\cal A}$. Then calculate the jumps: 
\[
\Delta_i(\widehat{\tau})=\widehat{\tau}^{(i+1)}-\widehat{\tau}^{(i)},\ \ i=1,\cdots,n-1.
\] 
If the number of distinct break points is assumed to be $K$, define $\overline{\tau}_j=\widehat{\tau}^{(i+1)}$ with $\Delta_i(\widehat{\tau})$ being the $j$-th largest jump, $1\leq j\leq K-1$. Then we sort $\overline{\tau}_1,\cdots,\overline{\tau}_{K-1}$ from minimum to maximum, denote them as $\overline{\tau}_K^{(1)},\cdots,\overline{\tau}_K^{(K-1)}$, and obtain the estimated clusters as 
\begin{equation}\label{eq4.7}
\widehat{\mathcal C}(k|K)=\left\{i\in\widehat{\mathcal C}_\bullet:\ \overline{\tau}_K^{(k-1)}\leq\widehat{\tau}_i<\overline{\tau}_K^{(k)} \right\},\ \ k=1,\cdots,K,
\end{equation}
where, without loss of generality, $\overline{\tau}_K^{(0)}=1$ and $\overline{\tau}_K^{(K)}=T$. Given the cluster number $K$, we compute 
\begin{equation}\label{eq4.8}
\widehat{\tau}_{k|K}=\frac{1}{\big|\widehat{\mathcal C}(k|K)\big|}\sum_{i\in\widehat{\mathcal C}(k|K)}\widehat{\tau}_i,\ \ \ \widehat{\nu}_{it,k|K}=\widehat{\mu}_{i,k|K}+\widehat{\delta}_{i,k|K}I\left(t>\widehat{\tau}_{k|K}\right)
\end{equation}
with
\[
\widehat{\mu}_{i,k|K}=\frac{1}{\widehat{\tau}_{k|K}}\sum_{t=1}^{\widehat{\tau}_{k|K}}X_{it}\ \ \ {\rm and}\ \ \ \widehat{\delta}_{i,k|K}=\frac{1}{T-\widehat{\tau}_{k|K}}\sum_{t=\widehat{\tau}_{k|K}+1}^TX_{it}-\frac{1}{\widehat{\tau}_{k|K}}\sum_{t=1}^{\widehat{\tau}_{k|K}}X_{it}.
\]
Construct the penalised objective function:
\begin{equation}\label{eq4.9}
{\sf IC}(K)=\ln(V(K))+K\rho_{NT},
\end{equation}
where $\rho_{NT}$ is a user-specified tuning parameter satisfying some restrictions, see Assumption~\ref{ass:2}(iii) below, and 
\[
V(K)= \frac{1}{\big|\widehat{\mathcal C}_\bullet\big|}\sum_{k=1}^K\sum_{i\in\widehat{\mathcal C}(k|K)}\frac{1}{T}\sum_{t=1}^T\left\Vert X_{it}-\widehat{\nu}_{it,k|K}\right\Vert^2.
\]
The true cluster number $K_0$ is determined by 
\begin{equation}\label{eq4.10}
\widehat{K}=\argmin_{1\leq K\leq \overline{K}} {\sf IC}(K),
\end{equation}
where $\overline{K}$ is a pre-specified upper bound of the cluster number. Replacing $K$ by $\widehat{K}$ in~\eqref{eq4.7}, we obtain 
\[
\widehat{\mathcal C}(b_k)=\widehat{\mathcal C}\big(k|\widehat{K}\big),\quad k=1,\cdots,\widehat{K}.
\]
The following assumption is required to derive the consistency property for $\widehat{K}$ and $\widehat{\mathcal C}(b_k)$.  

\begin{assumption}\label{ass:2}

{\em (i) The cluster-specific break points satisfy that $b_k=c_kT$ with $ 0<c_1<\cdots<c_{K_0}<1$,
and the latent cluster size satisfies that $\left\vert{\mathcal C}(b_k)\right\vert=d_k \left\vert {\mathcal C}_\bullet\right\vert$ with $d_k>0$ and $\sum_{k=1}^{K_0}d_k=1$.}

{\em (ii) Let $\Vert\mu_i\Vert$ and $\Vert\delta_i\Vert$ be bounded uniformly over $i$ and $\min_{i\in{\mathcal C}_\bullet}\Vert\delta_i\Vert$ be bounded away from zero.} 

{\em (iii) Let $\rho_{NT}$ satisfy $\rho_{NT}\rightarrow0$ and $T\rho_{NT}/[\ln(N\vee T)]^{1+\zeta}\rightarrow\infty$ for any $\zeta>0$.}

\end{assumption}

Assumption~\ref{ass:2}(i) indicates that the distance between distinct break points is of the same order as $T$ and the cardinality of ${\mathcal C}(b_k)$ is similar over $k=1,\cdots,K_0$. Assumption~\ref{ass:2}(ii) imposes some mild restrictions on sizes of $\Vert\mu_i\Vert$ and $\Vert\delta_i\Vert$. Finally, Assumption~\ref{ass:2}(iii) is a crucial condition on the tuning parameter in the penalty term, ensuring that the information criterion can consistently select $K_0$. 

\begin{theorem}\label{thm:3} Suppose that the latent structure~\eqref{eq4.6} and Assumptions~\ref{ass:1} and~\ref{ass:2} are satisfied. In addition, there exists $\kappa\geq0$ such that $N=O(T^{\kappa})$. Then we have
\begin{equation}\label{eq4.11}
{\sf P}\left(\widehat{K}=K_0\right)\rightarrow1,
\end{equation}
and
\begin{equation}\label{eq4.12}
{\sf P}\left(\widehat{\mathcal C}(b_k)={\mathcal C}(b_k):\ k=1,\cdots,K_0 \ |\ \widehat{K}=K_0\right)\rightarrow1.
\end{equation}

\end{theorem}

The above theorem establishes the consistency property for the proposed information criterion and cluster analysis of the heterogeneous break points. As $X_{it}$, $i\in{\mathcal C}(b_k)$, $t=1,\cdots,T$, have a common break point, it is sensible to estimate $b_k$ more efficiently by pooling the CUSUM quantities over the subjects in $\widehat{\mathcal C}(b_k)$, i.e.,
\begin{equation}\label{eq4.13}
\widehat{b}_k=\argmax_{1\leq t\leq T}\sum_{i\in\widehat{\mathcal C}(b_k)}\int_{\mathbb C} Z_{iT}^2(t/T;u)du,
\end{equation}
whose consistency property is given in the following theorem.

\begin{theorem}\label{thm:4}

Suppose that the latent structure~\eqref{eq4.6} and Assumptions~\ref{ass:1} and~\ref{ass:2}(i) are satisfied, $|{\mathcal C}_\bullet |=O(T^2)$, $T=O(|{\mathcal C}_\bullet|^{3/2})$, and
\begin{equation}\label{eq4.14}
\min_{1\leq k\leq K_0}\frac{1}{|{\mathcal C}(b_k)|^{1/2}}\sum_{i\in{\mathcal C}(b_k)}\Vert\delta_i\Vert^2\rightarrow\infty.
\end{equation}
In addition, $\eta_{it}$ are independent over $i$. Then, as $T$ and $\left\vert {\mathcal C}_\bullet\right\vert$ tend to infinity jointly,
\begin{equation}\label{eq4.15}
{\sf P}\left(\widehat{b}_k=b_k,\ k=1,\cdots,K_0\right)\rightarrow1.
\end{equation}

\end{theorem}

Theorem~\ref{thm:4} extends Theorem 3.1 in \cite{Ba10} to the high-dimensional functional data setting with a latent structure on the heterogeneous break points. The condition~\eqref{eq4.14} plays a key role in the theoretical derivation and is a natural extension of Assumption 2 in \cite{Ba10}. It indicates that break sizes cannot be too small over $i\in{\mathcal C}_\bullet$ so that the true break point can be consistently estimated. In particular, when $\min_{i\in{\mathcal C}_\bullet}\Vert\delta_i\Vert$ is bounded away from zero as in Assumption~\ref{ass:2}(ii),~\eqref{eq4.14} would be automatically satisfied if $|{\mathcal C}_\bullet|\rightarrow\infty$. Note that in the case of break estimation for a single time series process, the optimal asymptotic order for the break point estimation is $O_P(1)$ \citep[e.g.,][]{Ba97, ARS18}. With more sample information from pooling a panel of functional time series, it is unsurprising to improve the approximation order to $o_P(1)$. The restrictions $|{\mathcal C}_\bullet |=O(T^2)$ and $T=O(|{\mathcal C}_\bullet|^{3/2})$ indicate that $|{\mathcal C}_\bullet|$ may diverge to infinity at a faster rate than $T$. In addition, the cross-sectional independence condition on $\eta_{it}$ facilitates the technical proofs but may be replaced by some high-level conditions such as those in Assumption~\ref{ass:1}(ii).


\section{Monte-Carlo simulation}\label{sec5}
\renewcommand{\theequation}{5.\arabic{equation}}
\setcounter{equation}{0}


\subsection{Data generating process}\label{sec5.1}

Generate time series of random functions $[\epsilon_{i1}(u),\epsilon_{i2}(u),\cdots,\epsilon_{iT}(u)]$, $i=1,\cdots,N$, as follows
\begin{equation}\label{eq5.1}
\epsilon_{it}(u) = \sum_{j=1}^J\epsilon_{it,j}f_j(u),\ \  \epsilon_{it,j}=\beta_{it,j} + \eta_{it,j},\ \ u\in {\mathbb C}= [0,1]
\end{equation}
where $f_1(u)$, $f_2(u), \cdots, f_{J}(u)$ are randomly sampled without replacement from $J=21$ Fourier basis functions, and $\eta_{it,j}$ are innovations following ${\sf N}(0,1/j)$ independently over $i$, $t$ and $j$. To generate functional values, we consider $101$ equally-spaced grid points between $0$ and $1$. Writing $\boldsymbol{\beta}_{t,j} = \left(\beta_{1t,j},\beta_{2t,j},\cdots,\beta_{Nt,j}\right)^{^\intercal}$, $j=1,\cdots,21$, we generate $\boldsymbol{\beta}_{t,j}$ (independently over $j$) from a vector autoregression of order 1:
\begin{equation}\label{eq5.2}
\boldsymbol{\beta}_{t,j} = \boldsymbol{A}\boldsymbol{\beta}_{t-1,j} + \boldsymbol{z}_{t,j},
\end{equation}
where $\boldsymbol{A}=(a_{ij})_{N\times N}$ is the transition matrix, and $\boldsymbol{z}_{t,j}$ is independently generated by an $N$-dimensional normal distribution with mean zero and identity covariance matrix. Following \cite{LRS23}, $\boldsymbol{A}$ is a banded matrix with $a_{ij}$ independently generated from a ${\sf U}(-0.3,0.3)$ when $|i-j|\leq 3$ and $a_{ij}=0$ when $|i-j|>3$. In the simulation, we set $T=200$, $N=200$ or $400$, and the replication number $R=1000$. 

\smallskip

We randomly select ${\sf SDR} \times N$ subjects with a change point, where {\sf SDR} denotes a sparse-to-dense ratio. The remaining subjects have no change point. To specify a change-point location $\tau_i$ for the selected subject, we draw ${\sf SDR}\times N$ values from ${\sf U}(0.25 \times T, 0.75\times T)$. The lower and upper bounds of the uniform distribution are purposely chosen so that the change point location is not close to the boundary of a sample. As in \cite{ARS18}, for $i\in{\mathcal C}_\bullet$ (the index set for subjects with breaks), we define a class of break functions:
\[
\delta_{i,m}(u) = \delta_m^*(u)\times \sqrt{c_{i}^\ast},\ \ \delta_m^*(u)= \frac{1}{\sqrt{m}}\sum^m_{j=1}f_{j}(u), 
\]
where $m=1,\cdots,J$ and $c_i^\ast$ is a positive constant to be specified later. For each $i\in{\mathcal C}_\bullet$, $\delta_{i,1}(u)$ is the case of a break only in the leading eigendirection (determined by the basis function $f_1$), while $\delta_{i,J}(u)$ is the case of a break that affects all the eigendirections, see the discussion in \cite{ARS18}. The value $c_i^\ast$ controls the break magnitude, linking to the signal-to-noise ratio: 
\[
\text{SNR}_{i} = \frac{(\tau_i/T)(1-\tau_i/T)\|\delta_{i,m}\|^2}{\text{tr}({\boldsymbol\Omega}_{\epsilon_{i}})}=c_{i}^\ast\times\frac{(\tau_i/T)(1-\tau_i/T)}{\text{tr}({\boldsymbol\Omega}_{\epsilon_{i}})},
\]
where $\text{tr}(\cdot)$ denotes the trace of a square matrix and ${\boldsymbol\Omega}_{\epsilon_{i}}$ denotes the long-run covariance matrix of ${\boldsymbol\epsilon}_{it}=\left(\epsilon_{it,1},\epsilon_{it,2},\cdots, \epsilon_{it,21}\right)^{^\intercal}$ over the time span. The value of $c_{i}^\ast$ can be easily computed with a given SNR$_{i}$ level. For those subjects without breaks, we set $c_{i}^\ast=0$. With a given integer value of $m$, we finally simulate a panel of functional time series as follows,
\begin{equation}\label{eq5.3}
X_{it}(u) = \delta_{i,m}(u) \times I\left(t>\tau_{i}\right) + \epsilon_{it}(u).
\end{equation}

\subsection{Test results for structural breaks}\label{sec5.2}

In the simulation study, we consider the following two choices of high-criticism thresholds in the PE-CUSUM test statistic:
\begin{align*}
\xi_{NT,1} &= c_\xi\ln(NT)\times \ln\ln(NT), \\ 
\xi_{NT,2} &= c_\xi\ln(N\vee T) \ln\ln(N\vee T),
\end{align*}
where $c_\xi=\lambda_1^{1/2}$ with $\lambda_1$ being the leading eigenvalue of $\widetilde{\boldsymbol\Omega}$ in~\eqref{eq2.7}. The asymptotic theorems in Sections~\ref{sec3} and~\ref{sec4} are derived by setting $\xi_{NT}=\xi_{NT,2}$. In fact, they continue to hold with minor modifications when $\xi_{NT}=\xi_{NT,1}$. In the following simulation results, we denote these two PE-CUSUM test statistics as PE-CUSUM$_1$ and PE-CUSUM$_2$. We also consider the conventional CUSUM (without PE) defined in~\eqref{eq3.4} as a benchmark. We only report the simulation results when $m=1$ in the break function definition, i.e., breaks occur in the leading eigendirection, since the results are similar when $m$ is set as other positive integer values.

\smallskip

In Table~\ref{tab:1} below, we report the size performance of CUSUM and PE-CUSUM in finite samples, where three levels of significance $\alpha=0.01, 0.05$, and 0.10 are considered. The test critical values are determined by \cite{ARS18}'s simulation-based method. The sizes of all the three tests are generally close to the nominal ones. In particular, the results confirm the validity of Theorem~\ref{thm:1}(i) and show that incorporating the PE component in the CUSUM test statistic does not lead to severe size distortion. 

\bigskip

\begin{center}
\tabcolsep 0.3in
{\small\begin{longtable}{@{}llccc@{}}
\caption{Size performance of the CUSUM and PE-CUSUM tests}\label{tab:1} \\
\toprule
$N$ & test statistics & $\alpha=0.01$ & $\alpha=0.05$ & $\alpha=0.10$ \\
\hline
\endfirsthead
\toprule
$N$ & test statistics & $\alpha=0.01$ & $\alpha=0.05$ & $\alpha=0.10$ \\
\hline
\endhead
\hline \multicolumn{5}{r}{{Continued on next page}} \\
\endfoot
\endlastfoot
 200 	& PE-CUSUM$_1$ 	& 0.010 & 0.062 & 0.122 \\
		&PE-CUSUM$_2$ 	& 0.016 & 0.067 & 0.126 \\ 
		& CUSUM 	& 0.009 & 0.061 & 0.121  \\
		\\
 400 	&  PE-CUSUM$_1$	& 0.012 & 0.062 & 0.127 \\ 
		& PE-CUSUM$_2$ 	& 0.014 & 0.064 & 0.129 \\ 
		& CUSUM	& 0.012 & 0.062 & 0.127  \\
\bottomrule
\end{longtable}}
\end{center}

In Table~\ref{tab:3}, we report the power performance of the CUSUM and PE-CUSUM test statistics when SNRs are $10^{-1}, 10^{-2}$ and $5\times 10^{-3}$ and SDRs are $0.1$ and $0.5$. We note that percentages of rejecting the null hypothesis via the three test statistics are the same and close to $1$ when SDR is $0.5$, whereas both the PE-CUSUM$_1$ and PE-CUSUM$_2$ outperform the CUSUM statistic (without PE) when breaks are sparse (i.e., SDR is $0.1$) and the signal-to-noise ratios are low (i.e., SNRs are $10^{-2}$ and $5\times 10^{-3}$). The latter confirms the improvement of power performance with the extra PE component in the test statistic~\eqref{eq3.7}. Meanwhile, the performance of PE-CUSUM$_2$ is more stable than PE-CUSUM$_1$ in particular when SNR is $5\times 10^{-3}$, indicating that $\xi_{NT,2}= c_\xi\ln(N\vee T) \ln\ln(N\vee T)$ may be a more appropriate choice for the high-criticism threshold in PE-CUSUM.

{\small
\begin{center}
\tabcolsep 0.1in
\begin{longtable}{@{}lllcccccc@{}}	
\caption{Percentages of rejecting the null hypothesis over $1000$ replications}\label{tab:3}\\
\toprule
	&	& &	\multicolumn{3}{c}{SDR = 0.1} &  \multicolumn{3}{c}{SDR = 0.5}    \\
	\cline{2-9}
 SNR & $N$ & test statistics	& $\alpha=0.01$ & $\alpha=0.05$ & $\alpha=0.10$ & $\alpha=0.01$ & $\alpha=0.05$ & $\alpha=0.10$  \\\hline
\endfirsthead
\toprule
	&	& &	\multicolumn{3}{c}{SDR = 0.1} &	\multicolumn{3}{c}{SDR = 0.5}  \\
	\cline{2-9}
 SNR & $N$ & test statistics	& $\alpha=0.01$ & $\alpha=0.05$ & $\alpha=0.10$ & $\alpha=0.01$ & $\alpha=0.05$ & $\alpha=0.10$  \\\hline
\endhead
\hline \multicolumn{9}{r}{{Continued on next page}} \\
\endfoot
\endlastfoot
 $10^{-1}$ & 200 & PE-CUSUM$_1$ & 0.957 & 0.957 & 0.957 & 0.958 & 0.961 & 0.962 \\
	       & 	& PE-CUSUM$_2$ & 0.957 & 0.957 & 0.957& 0.958 & 0.961 & 0.962 \\
	       & 	& CUSUM & 0.957 & 0.957 & 0.957 & 0.958 & 0.961 & 0.962 \\ 
\\
	       & 400 & PE-CUSUM$_1$ & 0.953 & 0.953 & 0.953 &  0.955 & 0.959 & 0.962 \\  
		       & 	 & PE-CUSUM$_2$ & 0.953 & 0.953 & 0.953 & 0.955 & 0.959 & 0.962 \\ 
	       & 	 & CUSUM & 0.953 & 0.953 & 0.953 & 0.955 & 0.959 & 0.962 \\ 
\\
 $10^{-2}$ 		& 200 & PE-CUSUM$_1$ & 0.957 & 0.957 & 0.957 & 0.958 & 0.961 & 0.962  \\	
		& 	  & PE-CUSUM$_2$ & 0.957 & 0.957 & 0.957 & 0.958 & 0.961 & 0.962  \\
		& 	  & CUSUM   & 0.487 & 0.865 & 0.941 & 0.958 & 0.961 & 0.962  \\	
\\
	& 400 & PE-CUSUM$_1$ & 0.953 & 0.953 & 0.953 & 0.955 & 0.959 & 0.962 \\
		& 	  & PE-CUSUM$_2$ & 0.953 & 0.953 & 0.953 & 0.955 & 0.959 & 0.962 \\
		& 	  & CUSUM   & 0.953 & 0.953 & 0.953 & 0.955 & 0.959 & 0.962 \\ 
\\
$5\times 10^{-3}$ & 200 & PE-CUSUM$_1$ & 0.685 & 0.686 & 0.686 & 0.957 & 0.957 & 0.957  \\
		& 	& PE-CUSUM$_2$ & 0.957 & 0.957 & 0.957 & 0.957 & 0.957 & 0.957 \\
		& 	& CUSUM    & 0.018 & 0.041 & 0.047 & 0.957 & 0.957 & 0.957 \\
\\	
		& 400 & PE-CUSUM$_1$ & 0.768 & 0.914 & 0.948 & 0.953 & 0.953 & 0.953 \\
		& 	  & PE-CUSUM$_2$ & 0.953 & 0.953 & 0.953 & 0.953 & 0.953 & 0.953  \\
	& 	  & CUSUM & 0.444 & 0.882 & 0.941 & 0.953 & 0.953 & 0.953 \\
	
\bottomrule
\end{longtable}
\end{center}
}

\subsection{Estimation results of subjects with breaks}\label{sec5.3}

We next compute the percentage of correctly identifying subjects with a change point, i.e, ${\mathcal C}_\bullet$ defined in Section~\ref{sec4}. It is defined by $\text{TP}/N$, where TP denotes the number of true positive hits (the subject indices match with the estimated indices), and $N$ denotes the number of subjects. Let the subjects with breaks be categorised as one, and those without breaks be categorised as zero. We also compute the $F_1$ score, which is the harmonic mean of precision defined by
\[
\text{F}_1 = \frac{\text{TP}}{\text{TP}+(\text{FP}+\text{FN})/2},
\]
where FP denotes false alarm or overestimation (type I error, where a change point is detected for subjects without breaks), and FN denotes miss or underestimation (type II error, where a change point is not detected for subjects with breaks). 

\smallskip

The TP and $\text{F}_1$ score results are reported in Table~\ref{tab:2}, when SNRs are $10^{-1}, 10^{-2}$ and $5\times 10^{-3}$ and SDRs are $0.1$ and $0.5$. As SNR decreases from $10^{-1}$ to $5\times 10^{-3}$, there is a decrease of the TP and F$_1$ values. This is unsurprising as it becomes more difficult to detect breaks when the signal-to-noise ratio is lower. The estimation $\widehat{\mathcal C}_\bullet$ using $\xi_{NT}=\xi_{NT,2}$ performs better than that with $\xi_{NT}=\xi_{NT,1}$, which is consistent with the discussion in Section~\ref{sec5.2}, and again justifies the use of $\xi_{NT}= c_\xi\ln(N\vee T) \ln\ln(N\vee T)$ as the high-criticism threshold.


\begin{table}[!htbp]
\centering
\caption{The TP and F$_1$ measurements of correctly identifying ${\mathcal C}_\bullet$}\label{tab:2}
\tabcolsep 0.155in
\begin{tabular}{@{}ccccccc@{}}
\toprule
&	& 	& \multicolumn{2}{c}{TP} & \multicolumn{2}{c}{$\text{F}_1$ Score}  \\ 
SDR & SNR &  $N$ & $\xi_{NT,1}$ & $\xi_{NT,2}$ & $\xi_{NT,1}$ & $\xi_{NT,2}$ \\
\midrule
0.1 	& $10^{-1}$ 	& 200 	& 1 & 1 & 1 & 1   \\
	&			& 400 	& 1 & 1 & 1 & 1   \\
\\
	& $10^{-2}$ 	& 200 & 0.972 & 1 & 0.985 & 1  \\
	&			& 400 & 0.962 & 1 & 0.979 & 1 \\
\\
	& $5\times 10^{-3}$ 	& 200 & 0.910 & 0.997 & 0.952 & 0.999 \\
	&				& 400 & 0.906 & 0.988 & 0.950 & 0.994 \\
				\\
0.5& $10^{-1}$ 	& 200 &  1 & 1 & 1 & 1  \\
	&		& 400 &  1 & 1 & 1 & 1  \\
\\
	& $10^{-2}$ 	& 200 & 0.861 & 1 & 0.880 & 1 \\
	&			& 400 & 0.808 & 1 & 0.841 & 1 \\
\\
	& $5\times 10^{-3}$ 	& 200 & 0.548 & 0.986 & 0.690 & 0.986 \\
	&				& 400 & 0.527 & 0.940 & 0.680 & 0.944 \\
\bottomrule
\end{tabular}
\end{table}

\subsection{Estimation results of latent groups for change points}\label{sec5.4}

We next assess the clustering methodology proposed in Section~\ref{sec4} by imposing a latent group structure on change points. Specifically, we set the number of distinct change points as $K_0=3$ and split the $N$ subjects into the following four groups: the first half of $N$ subjects have no change point; and the remaining subjects are equally split into three groups with the change point at $0.25\times T$, $0.5\times T$ and $0.75\times T$, respectively. The tuning parameter used in the information criterion is set as $\rho_{NT}=(N\vee T)^{-1/2} \ln(N \vee T)$, so that Assumption~\ref{ass:2}(iii) is satisfied (if $N=T^\phi$ with $\phi<2$). To assess the estimation accuracy of group membership, we compute the Purity and normalised mutual information (NMI) measurements which are respectively defined as
\[
\text{Purity}\left(\widehat{\mathcal C}_\bullet, {\mathcal C}_\bullet\right) = \frac{1}{N}\sum^{\widehat{K}}_{k=1}\max_{1\leq j\leq K_0}\left|\widehat{\mathcal{\mathcal C}}_k\cap {\mathcal C}_j\right|,
\]
and
\[
\text{NMI}\left(\widehat{\mathcal C}_\bullet, {\mathcal C}_\bullet\right) = 2\frac{ I\left(\widehat{C}_\bullet, {\mathcal C}_\bullet\right)}{H(\widehat{\mathcal C}_\bullet)+H({\mathcal C}_\bullet)},
\]
where $\widehat{\mathcal C}_\bullet = \left\{\mathcal{\mathcal C}_1,\cdots,\mathcal{\mathcal C}_{\widehat{K}}\right\}$ is the estimate of ${\mathcal C}_\bullet=\left\{\mathcal{C}_1,\cdots, \mathcal{C}_{K_0}\right\}$, $H({\mathcal C}_\bullet)$ denotes the entropy of ${\mathcal C}_\bullet$, $I\left(\widehat{\mathcal C}_\bullet, {\mathcal C}_\bullet\right)$ is the mutual information between $\widehat{\mathcal C}_\bullet$ and ${\mathcal C}_\bullet$ defined by
\[
I\left(\widehat{\mathcal C}_\bullet, {\mathcal C}_\bullet\right) = \sum^{\widehat{K}}_{k=1}\sum^{K_0}_{j=1}\left(\frac{|\widehat{\mathcal{C}}_k \cap \mathcal{C}_j|}{N}\right)\log_2\left(\frac{N|\widehat{\mathcal{C}}_k\cap {\mathbb C}_j|}{|\widehat{\mathcal{C}}_k||\mathcal{C}_j|}\right).
\]

\smallskip

The relevant results are summarised in Tables~\ref{tab:4.1} and~\ref{tab:4.2}. It follows from Table~\ref{tab:4.1} that the frequency of correctly estimating the cluster number generally decreases as the SNR decreases from $10^{-1}$ to $5\times 10^{-3}$ (because it becomes more difficult to detect breaks when SNR is lower). The information criterion with $\xi_{NT}=\xi_{NT,2}$ can more accurately estimate the cluster number than that with $\xi_{NT}=\xi_{NT,1}$. In particular, when SNRs are $10^{-2}$ and $5\times 10^{-3}$, the information criterion with $\xi_{NT}=\xi_{NT,1}$ tends to under-estimate the cluster number. The Purity and NMI values in Table~\ref{tab:4.2} are close to the perfect value of one when SNR is $10^{-1}$, and gradually decrease when SNR becomes smaller.



\begin{table}[!htb]
\centering
\tabcolsep 0.25in
\caption{Percentages of accurately estimating the cluster number}\label{tab:4.1}
\begin{tabular}{@{}llccc@{}}	
\toprule
$\xi_{NT}$	 &N & SNR=$10^{-1}$ & SNR=$10^{-2}$ & SNR=$5\times10^{-3}$ \\
\midrule
$\xi_{NT,1}$ & 200  & 0.875 &   0.518  &   0.099    \\
$\xi_{NT,1}$ & 400  & 0.807   & 0.512 &   0.064 \\
$\xi_{NT,2}$ & 200  & 0.955   & 0.879   & 0.861  \\
$\xi_{NT,2}$  & 400  & 0.953  & 0.895   & 0.789  \\
\bottomrule		
\end{tabular}
\end{table}



\begin{table}[!htb]
\centering
\tabcolsep 0.2in
\caption{The Purity and NMI measurements for cluster membership estimation}\label{tab:4.2}
\begin{tabular}{@{}llcccc@{}}	
\toprule
SNR & $N$ & Purity ($\xi_{NT,1}$) & Purity ($\xi_{NT,2}$) & NMI ($\xi_{NT,1}$) & NMI ($\xi_{NT,2}$) \\
\midrule
$10^{-1}$  	& 200 & 0.986 & 1 & 0.984 & 1 \\
			& 400 & 0.975 & 1 & 0.972 & 1 \\
\\
$10^{-2}$ 		& 200  & 0.913 & 0.987 & 0.878 & 0.996 \\
			& 400 & 0.914 & 0.991 & 0.881 & 0.997 \\
\\
$5\times 10^{-3}$  & 200 & 0.744 & 0.981 & 0.533 & 0.956 \\
			     & 400  & 0.734 & 0.969 & 0.522 & 0.902 \\
\bottomrule		
\end{tabular}
\end{table}

We finally assess the performance of the estimated break locations, and compare the post-clustering estimation~\eqref{eq4.13} with the pre-clustering estimation~\eqref{eq4.1}, from which we may demonstrate the usefulness of clustering heterogeneous change points over subjects. We compute the mean squared distances between the estimated and true change points with the results reported in Table~\ref{tab:5}. According to the previous simulation results, it may be more appropriate to use $\xi_{NT}=\xi_{NT,2}$ in the break location estimation. Table~\ref{tab:5} shows that, conditional on the accurate estimation of the cluster number, the post-clustering estimation performs significantly better than the pre-clustering one which ignores the latent group structure on the heterogeneous change points. This is consistent with the convergence results in Theorems~\ref{thm:2} and~\ref{thm:4}, which state that the post-clustering estimation is consistent, whereas the pre-clustering estimation has the logarithmic-($N\vee T$) approximation order.



\begin{table}[!htb]
\centering
\tabcolsep 0.3in
\caption{\small Mean squared distances between the estimated and true change points (conditional on $\widehat{K}=3$)}\label{tab:5}
\begin{tabular}{@{}llccc@{}}
\toprule
Estimation	 &	$N$ &  SNR=$10^{-1}$ & SNR=$10^{-2}$ & SNR=$5\times10^{-3}$ \\
\midrule
Post-clustering & 200 &  0  & 0.234  & 0.910 \\
Post-clustering & 400 &  0   & 0.199  & 0.753 \\
Pre-clustering  & 200 & 0.002   & 1.007  & 4.591 	\\
Pre-clustering  & 400 & 0.002 &  0.982 &  4.477	\\
\bottomrule
\end{tabular}
\end{table}


\section{Empirical applications}\label{sec6}
\renewcommand{\theequation}{6.\arabic{equation}}
\setcounter{equation}{0}

In this section, we apply the developed break detection and estimation methods to two empirical data sets: $28$ Dow Jones Industrial Average (DJIA) constituent stocks from January 2, 2018 to December 31, 2021, and the age-specific mortality rates collected in $32$ countries from 1960 to 2013.

\subsection{DJIA and its constituent stocks}\label{sec:6.1}

The DJIA index shows how $30$ publicly owned large companies based in the United States have traded during a standard New York Stock Exchange trading session. Table~\ref{tab:DJIA} lists the stock names and tick symbols of 30 constituents for the DJIA index. We consider their daily cross-sectional returns from January 2, 2018 to December 31, 2021, with the data obtained from the Refinitiv Datascope (\url{https://select.datascope.refinitiv.com/DataScope/}). There are $T=1,008$ trading days. Among the 30 constituent stocks, DOW and HON.O began trading on April 2, 2019 and May 11, 2021, respectively. Thus, we remove these two stocks in our empirical analysis, resulting in $N=28$. For each trading day, we consider $5$-minute resolution data covering the period between 9:30 and 15:55 Eastern standard time, and obtain $78$ data points. For asset $i$, let $P_{it}(u_j)$ be the intraday 5-minute close price at time $u_j$ on trading day $t$, and construct a sequence of CIDRs \citep[e.g.,][]{RWZ20}:
\[
X_{it}(u_j) = 100\times [\ln P_{it}(u_j) - \ln P_{it}(u_1)], \qquad j=2,3,\cdots,78, 
\]
where $i=1,2,\cdots,28$ and $t=1,\cdots,1,008$. We use the linear interpolation algorithm \citep{Hyndman19} to convert discrete data points into a continuous function.

\begin{table}[!htb]
\centering
\caption{Stock names and tick symbols of 30 constituents of the DJIA index.}\label{tab:DJIA}
\tabcolsep 0.28in
{\small\begin{tabular}{@{}llll@{}}
\toprule
Tick symbol  & Stock name  & Tick symbol & Stock name \\
\midrule
AAPL & Apple & JNJ & Johnson \& Johnson \\
AMGN & Amgen & JPM & JPMorgan Chase \\
AXP & American Express & KO & Coca-Cola \\
BA & Boeing & MCD & McDonald's \\
CAT & Caterpillar & MMM & 3M \\
CRM & Salesforce & MRK & Merck \\
CSCO & Cisco  & MSFT & Microsoft \\
CVX & Chevron & NKE & Nike \\
DIS & Disney & PG & Procter \& Gamble \\
DOW & Dow Chemical & TRV & Travelers Companies \\
GS & Goldman Sachs & UNH & United Health \\
HD & Home Depot & V & Visa \\
HON & Honeywell & VZ & Verizon \\
IBM & International Business Machines & WBA & Walgreen \\
INTC & Intel & WMT & Wal-Mart \\
\bottomrule
\end{tabular}}
\end{table}

We implement the proposed PE-CUSUM test to detect if there exists a structural break in at least one of the $28$ constituents. The test $p$-value is 0, indicating the existence of breaks. We then estimate the heterogeneous change points using~\eqref{eq4.1} and identify the constituents with a structural break. The estimation results are reported in Table~\ref{tab:DJIA_change_point}. From the estimated break dates, it is sensible to expect a common change point (likely in March 2023) shared by some stocks. Hence, we further implement the clustering algorithm proposed in Section~\ref{sec4} to estimate the number of clusters and common break date. The information criterion~\eqref{eq4.10} selects the number of clusters as one, and the post-clustering pooled CUSUM method in~\eqref{eq4.13} estimates the common break date on March 16, 2020. This is consistent with our observation of the heterogeneous break date estimates in Table~\ref{tab:DJIA_change_point}. The common break is related to the stock market crash in March 2020 when the DJIA index suffered severe losses on March 9 (-7.79\%), March 12 (-9.99\%), and March 16 (-12.93\%).

\begin{table}[!htb]
\centering
\tabcolsep 0.55in
\caption{The estimated heterogeneous change points in the $22$ companies traded in the DJIA index.}\label{tab:DJIA_change_point}
{\small\begin{tabular}{@{}llll@{}}
\toprule
Tick symbol  & Change point & Tick symbol   & Change point \\  
\midrule
AAPL 	& 2021-03-08 & IBM 		& 2020-03-13 \\ 
AMGN 	& 2020-02-28 & JPM 	& 2019-03-29 \\ 
AXP 		& 2020-03-23 & MCD 	& 2020-03-18 \\ 
BA 		& 2019-11-15 & MSFT 	& 2019-11-06 \\ 
CAT 		& 2018-12-24 & NKE 	& 2020-03-13 \\ 
CRM 	& 2020-08-27 & PG 		& 2020-03-12 \\ 
CSCO 	& 2020-03-13 & TRV 	& 2020-03-18 \\ 
CVX 		& 2020-03-23 & UNH 	& 2020-03-23 \\ 
DIS 		& 2020-03-18 & V 		& 2020-08-28 \\ 
GS 		& 2018-12-26 & WBA 	& 2020-09-10 \\ 
HD 		& 2020-03-18 & WMT 	& 2020-02-28 \\ 
\bottomrule
\end{tabular}}
\end{table}

\subsection{Multi-country age-specific mortality rates}\label{sec:6.2}

The age-specific mortality rates are obtained from \cite{HMD23}. Our dataset covers the period from 1960 to 2013 and $32$ countries with sufficient data to use. Table~\ref{tab:mortality_list} shows a list of these countries and the corresponding ISO Alpha-3 codes. For each of the 32 countries, we consider the ages from 0 to 99 in a calendar year and the last age group 100+. We smooth the age-specific mortality rates by a weighted penalised regression with monotonic constraint \citep[e.g.,][]{Wood94, HU07}. The same data set is also considered by \cite{TSY22}. 

\begin{table}[!htb]
\tabcolsep 0.1in
\centering
\caption{List of selected countries and corresponded ISO Alpha-3 codes.}\label{tab:mortality_list}
{\small\begin{tabular}{@{}llllllll@{}}
\toprule
Country & Code & Country & Code & Country & Code & Country & Code \\
\midrule
Australia & AUS & Estonia & EST & Lithuania & LTU & Russia & RUS \\
Austria & AUT & Finland & FIN & Latvia & LVA & Slovakia & SVK \\
Belgium & BEL & France & FRA & Luxembourg & LUX & Spain & ESP \\
Belarus & BLR & Hungary & HUN & Norway & NOR & Sweden & SWE \\
Bulgaria & BGR & Iceland & ISL & Portugal & PRT & Switzerland & CHE \\
Canada & CAN & Ireland & IRE & Poland & POL & Great Britain & GBR \\
Denmark & DNK & Italy & ITA & Netherlands & NLD & United States & USA \\
Czech Republic & CZE & Japan & JPN & New Zealand & NZL & Ukraine & UKR \\
\bottomrule
\end{tabular}}
\end{table}

The developed PE-CUSUM test rejects the null hypothesis of no structural break. For the female population, Japan is the only country with a structural break and the estimated change point is 1983. For the male population, we use~\eqref{eq4.1} to estimate the heterogeneous change points and identify the countries with a structural break. The estimation results are reported in Table~\ref{tab:mort_country}, where we find that many developed countries experienced a structural break. It seems that a common break may occur in the late 1980s for those countries. This is confirmed by implementing the proposed clustering algorithm: the information criterion determines one cluster and the post-clustering CUSUM estimate of the common change point is 1988. 

\begin{table}[!htb]
\tabcolsep 0.13in
\centering
\caption{Estimated change points in 10 out of 32 countries (for the male population).}\label{tab:mort_country}
{\small\begin{tabular}{@{}lllllllllll@{}}
\toprule
 & \multicolumn{10}{c}{Country} \\
\midrule
Country 	& AUS	& AUT	& CAN	& CHE	& FIN	& FRA	& ITA	& JPN	& NZL	& PRT \\
Change point	& 1989	& 1986	& 1987	& 1991	& 1985	& 1989	& 1986	&1983	&1990	& 1991 \\
\bottomrule 
\end{tabular}}
\end{table}


\section{Conclusion}\label{sec7}

We propose a new fully functional test statistic combining the classic CUSUM and an extra PE component to detect structural breaks in the heterogeneous mean functions for large-scale functional time series, where the number of subjects may be larger than the time series length. The underlying functional time series are allowed to be weakly correlated over subjects. We derive the asymptotic property for the developed test under both the null and alternative hypotheses. In particular, the involvement of the PE component in the test statistic can enlarge the region where the test has power and detect sparse breaks in the alternative. We further impose a latent group structure on the heterogeneous break points and combine a simple clustering algorithm with an information criterion to accurately estimate the group membership and number. The post-clustering pooled CUSUM method using the estimated group structure is introduced to consistently locate the homogenous break point within each group. The Monte-Carlo simulation studies demonstrate the power enhancement property of the developed test in finite samples when breaks are sparse, the accuracy of the latent group structure estimation, and the convergence improvement of the post-clustering break point estimation (over the estimation neglecting the latent structure). The developed methodology is applied to detect and estimate structural breaks for DJIA constituent stocks and the age-specific mortality rates collected in $32$ countries.


\section*{Acknowledgements}

The first author is partially supported by the Australian Research Council Discovery Project (DP230102250) and the National Natural Science Foundation of China (72033002). The second author is partially supported by the National Science Foundation grant (DMS 1820702) and National Institutes of Health grants (R01AI136664 and R01AI170249). The third author is partially supported by the Australian Research Council Discovery Project (DP230102250). The usual disclaimer applies.

\bigskip
\bigskip
\bigskip


\appendix
\renewcommand{\theequation}{A.\arabic{equation}}
\setcounter{equation}{0}

\noindent{\Large\bf Appendix A:\ Proofs of the main asymptotic results}

\bigskip

\noindent In this appendix, we provide proofs of the main asymptotic theorems in Sections~\ref{sec3} and~\ref{sec4}. We start with some propositions whose proofs are available in a supplemental document.

\renewcommand{\theprop}{A.\arabic{prop}}
\setcounter{prop}{0}

\begin{prop}\label{prop:A.1}

 {\em Suppose that the conditions of Theorem~\ref{thm:1} are satisfied. Let 
\[\widetilde{S}_{NT}^\epsilon(x;u)=N^{1/2}\sum_{s=1}^{\lfloor Tx\rfloor}\widetilde{\epsilon}_{s}(u)\ \ {\rm and}\ \ G(x;u)=\sum_{i=1}^\infty\lambda_i^{1/2}W_i(x)\psi_i(u),\] 
where $\lambda_i$ and $\psi_i(\cdot)$ are defined as in Section~\ref{sec2} and $W_i(\cdot)$ are independent standard Brownian motions. There exists a sequence of Gaussian processes $\left\{G_{NT}(x;u):\ 0\leq x\leq 1,\ u\in{\mathbb C}\right\}$ whose probability measure converges weakly to that of $\left\{G(x;u):\ 0\leq x\leq 1,\ u\in{\mathbb C}\right\}$, and}
\begin{equation}\label{eqA.1}
\sup_{0\leq x\leq 1}\int _{\mathbb C} \left[\frac{1}{\sqrt{T}}\widetilde{S}_{NT}^\epsilon(x;u)-G_{NT}(x;u)\right]^2du=o_P(1).
\end{equation}

\end{prop}

\begin{prop}\label{prop:A.2}

 {\em Suppose that the conditions of Theorem~\ref{thm:1} are satisfied. Then we have
\begin{equation}\label{eqA.2}
{\sf P}\left(\max_{1\leq i\leq N}\sup_{0\leq x\leq 1} \int_{\mathbb C} Z_{iT}^2(x;u)du>\xi_{NT}\ |\ H_0\right)\rightarrow0,
\end{equation}
where $Z_{iT}(x;u)$ is defined in~\eqref{eq3.5} and $\xi_{NT}$ is defined as in Section~\ref{sec3}.}

\end{prop}

\begin{prop}\label{prop:A.3}

{\em Suppose that the conditions of Theorem~\ref{thm:4} are satisfied. Write
\[Z(t; {\mathcal C}(b_k))=\sum_{i\in{\mathcal C}(b_k)}\int_{\mathbb C} Z_{iT}^2(t/T;u)du,\ \ k=1,\cdots,K_0.\]
Then we have
\begin{equation}\label{eqA.3}
\max_{1\leq t\leq T}\left\vert Z(t; {\mathcal C}(b_k))-{\sf E}[ Z(t; {\mathcal C}(b_k))]\right\vert =O_P\left(|{\mathcal C}(b_k)|^{1/2}+\left(T\sum_{i\in{\mathcal C}(b_k)}\Vert\delta_i\Vert^2\right)^{1/2}\right)
\end{equation}
for $k=1,\cdots,K_0$, under $H_A$. }

\end{prop}

\begin{prop}\label{prop:A.4}

{\em Suppose that the conditions of Theorem~\ref{thm:4} are satisfied. For any $k=1,\cdots,K_0$ and $1\leq t\leq T$
\begin{equation}\label{eqA.4}
 {\sf E}\left[Z(b_k; {\mathcal C}(b_k))\right]-{\sf E} \left[Z(t; {\mathcal C}(b_k))\right]\geq m_0|t-b_k|\sum_{i\in{\mathcal C}(b_k)}\Vert\delta_i\Vert^2
\end{equation}
under $H_A$, where $m_0$ is a positive constant.}

\end{prop}

\begin{prop}\label{prop:A.5}

{\em Suppose that the conditions of Theorem~\ref{thm:4} are satisfied. Then we have
\begin{equation}\label{eqA.5}
{\sf P}\left(\max_{t\in {\cal N}_k(\epsilon)\backslash b_k}Z(t; {\mathcal C}(b_k))< Z(b_k; {\mathcal C}(b_k))\right)\rightarrow1,\ \ k=1,\cdots,K_0,
\end{equation}
under $H_A$, where ${\cal N}_k(\epsilon)=\left\{1\leq t\leq T:\ |t-b_k|\leq \varepsilon\varpi_k\right\}$ for any $\varepsilon>0$ and $\varpi_k=T^{1/2}|{\mathcal C}(b_k)|^{-1/4}$.}

\end{prop}

\noindent{\bf Proof of Theorem~\ref{thm:1}}.\ \ (i)\ Note that 
\[
\widetilde{Z}_{NT}(x;u)=\widetilde{Z}_{NT}^\epsilon(x;u):=\frac{1}{\sqrt{T}}\left[\widetilde{S}_{NT}^\epsilon(x;u)-\frac{\lfloor Tx\rfloor}{T}\widetilde{S}_{NT}^\epsilon(1;u)\right]
\]
under $H_0$, where $\widetilde{S}_{NT}^\epsilon(x;u)$ is defined in Proposition~\ref{prop:A.1}. By Proposition~\ref{prop:A.1} and the continuous mapping theorem \citep[e.g.,][]{Bi68}, we can prove that
\begin{equation}\label{eqA.6}
Z_{NT}\stackrel{d}\longrightarrow\sup_{0\leq x\leq 1}\sum_{i=1}^\infty\lambda_iB_i^2(x),
\end{equation}
which, together with Proposition~\ref{prop:A.2}, leads to~\eqref{eq3.10}.

\smallskip

(ii)\ Note that, under $H_A^\diamond$, there must exist an $i_0$ so that $\frac{T\omega_{Ti_0}^2\Vert\delta_{i_0}\Vert^2}{\xi_{NT}}\rightarrow\infty$. 
Let
\[
Z_{iT}^\epsilon(x;u)=\frac{1}{\sqrt{T}}\left[S_{iT}^\epsilon(x;u)-\frac{t}{T}S_{iT}^\epsilon(1;u)\right]\ \ {\rm with}\ \ S_{iT}^\epsilon(x;u)=\sum_{s=1}^{\lfloor Tx\rfloor} \epsilon_{is}(u),
\]
and write
\begin{equation}\label{eqA.7}
Z_{i_0T}(x_{i_0},u)=Z_{i_0T}^\epsilon(x_{i_0},u)-\sqrt{T}x_{i_0}(1-x_{i_0})\delta_{i_0}(u),
\end{equation}
where $x_{i_0}=\tau_{i_0}/T$. By Proposition~\ref{prop:A.2}, we have
\begin{equation}\label{eqA.8}
\max_{1\leq i\leq N}\sup_{0\leq x\leq 1}\int_{\mathbb C}\left[Z_{iT}^\epsilon(x;u)\right]^2du=o_P\left(\xi_{NT}\right).
\end{equation}
On the other hand, it is easy to verify that
\begin{equation}\label{eqA.9}
\frac{T\left[x_{i_0}(1-x_{i_0})\right]^2\Vert\delta_{i_0}\Vert^2}{\xi_{NT}}\geq \frac{T\omega_{Ti_0}^2\Vert\delta_{i_0}\Vert^2}{4\xi_{NT}}\rightarrow\infty.
\end{equation}
With~\eqref{eqA.7}--\eqref{eqA.9}, we may show that
\[
\sup_{0\leq x\leq 1}\int_{\mathbb C} Z_{i_0T}^2(x;u)du\geq \int_{\mathbb C} Z_{i_0T}^2(x_{i_0};u)du>\xi_{NT},\ \ {\it w.p.a.1},
\]
which, together with the definition of $Z_{NT}^\diamond$, leads to~\eqref{eq3.11}. 

Observe that
\[
\widetilde{Z}_{NT}(x;u)=\widetilde{Z}_{NT}^\epsilon(x;u)+\nu_{NT}(x;u),
\]
where $\nu_{NT}(x;u)$ is defined in~\eqref{eq3.9}. Hence, we have
\begin{equation}\label{eqA.10}
\int_{\mathbb C}\widetilde{Z}_{NT}^2(x;u)du=\int_{\mathbb C}\left[\widetilde{Z}_{NT}^\epsilon(x;u)\right]^2du+\int_{\mathbb C}\nu_{NT}^2(x;u)du+2\int_{\mathbb C}\widetilde{Z}_{NT}^\epsilon(x;u)\nu_{NT}(x;u)du.
\end{equation}
By~\eqref{eq3.9}, (\ref{eq3.10}), (\ref{eqA.10}) and the Cauchy-Schwarz inequality, we may show that $\sup_{0\leq x\leq 1}\int_{\mathbb C}\nu_{NT}^2(x;u)du$ is the asymptotic leading term of $Z_{NT}$ under $H_A$. Therefore, if $\nu_{NT}$ satisfies~\eqref{eq3.9}, we have  
\begin{equation}\label{eqA.11}
{\sf P}\left(Z_{NT}\geq z_\alpha\right)\rightarrow1.
\end{equation} 
A combination of~\eqref{eq3.11} and~\eqref{eqA.11} leads to~\eqref{eq3.12}. \hfill$\Box$ 

\medskip

\noindent{\bf Proof of Theorem~\ref{thm:2}}.\ \ By Proposition~\ref{prop:A.2}, we may show that 
\[
{\sf P}\left(\max_{i\in{\mathcal C}_\circ}\sup_{0\leq x\leq 1}\int_{\mathbb C} Z_{iT}^2(x;u)du\geq\xi_{NT}\right)\rightarrow0,
\]
which indicates that ${\sf P}\left({\mathcal C}_\circ\subset\widehat{\mathcal C}_\circ\right)\rightarrow1$. On the other hand, for any $i\notin{\mathcal C}_\circ$, using~\eqref{eq4.2} and following the argument in the proof of~\eqref{eq3.11}, we must have  
\[
{\sf P}\left(\min_{i\notin{\mathcal C}_\circ}\sup_{0\leq x\leq 1}\int_{\mathbb C} Z_{iT}^2(x;u)du>\xi_{NT}\right)={\sf P}\left(\min_{i\in{\mathcal C}_\bullet}\sup_{0\leq x\leq 1}\int_{\mathbb C} Z_{iT}^2(x;u)du>\xi_{NT}\right)\rightarrow1,
\]
and thus $i\notin\widehat{\mathcal C}_\circ$ {\em w.p.a.1}. Then we have ${\sf P}\left(\widehat{\mathcal C}_\circ\subset{\mathcal C}_\circ\right)\rightarrow1$. Combining the above arguments, we readily have that ${\sf P}\left(\widehat{\mathcal C}_\circ={\mathcal C}_\circ\right)\rightarrow1$. The proof of ${\sf P}\left(\widehat{\mathcal C}_\bullet={\mathcal C}_\bullet\right)\rightarrow1$ is analogous and thus skipped to save space. The proof of~\eqref{eq4.3} is completed.

\smallskip

We next turn to the proof of~\eqref{eq4.5}. Without loss of generality, we first consider the case of $1\leq t<\tau_i$ for $i\in{\mathcal C}_\bullet$. For notational simplicity, we write $Z_{it}=\left(Z_{iT}(t/T;u):\ u\in{\mathbb C}\right)$ and $\Vert Z_{it}\Vert^2=\int_{\mathbb C}Z_{iT}^2(t/T;u)du$. For $i\in{\mathcal C}_\bullet$, we observe that
\begin{equation}\label{eqA.12}
\Vert Z_{it}\Vert^2-\Vert Z_{i\tau_i}\Vert^2=\left\langle Z_{it}+Z_{i\tau_i},Z_{it}-Z_{i\tau_i}\right\rangle=\frac{1}{T}\left\langle P_{it,1}+Q_{it,1},P_{it,2}+Q_{it,2}\right\rangle,
\end{equation}
where
\begin{eqnarray}
P_{it,1}&=&2\sum_{s=1}^t\epsilon_{is}+\sum_{s=t+1}^{\tau_i}\epsilon_{is}-\frac{\tau_i+t}{T}\sum_{s=1}^T\epsilon_{is},\ \ \ P_{it,2}=-\sum_{s=t+1}^{\tau_i}\epsilon_{is}+\frac{\tau_i-t}{T}\sum_{s=1}^T\epsilon_{is},\notag\\
\ Q_{it,1}&=&-\frac{(T-\tau_i)(\tau_i+t)}{T}\delta_i,\ \ \ Q_{it,2}=\frac{(T-\tau_i)(\tau_i-t)}{T}\delta_i.\notag
\end{eqnarray}
Write
\[
Q_i(t)=\frac{1}{T}\left\langle Q_{it,1},Q_{it,2}\right\rangle=\frac{(t+\tau_i)(t-\tau_i)}{T}\left(\frac{T-\tau_i}{T}\right)^2\Vert\delta_i\Vert^2,
\]
which is an increasing function over $1\leq t\leq\tau_i$. By~\eqref{eq4.2}, we have
\begin{eqnarray}
\min_{i\in{\mathcal C}_\bullet}\max_{1\leq t\leq \tau_i-\theta[\ln(N\vee T)]^{1+\zeta}}\frac{1}{T}\left\langle Q_{it,1},Q_{it,2}\right\rangle&=&\min_{i\in{\mathcal C}_\bullet}Q_i\left(\lfloor \tau_i-\theta[\ln(N\vee T)]^{1+\zeta}\rfloor\right)\notag\\
&\rightarrow&-\theta\left[\frac{2\tau_i}{T}\cdot\left(\frac{T-\tau_i}{T}\right)^2\Vert\delta_i\Vert^2\right][\ln(N\vee T)]^{1+\zeta}\notag\\
&\leq&-\theta m_1 [\ln(N\vee T)]^{1+\zeta}\rightarrow-\infty,\label{eqA.13}
\end{eqnarray}
where $\theta$ is any positive number and $m_1$ is a positive constant. Note that $\left\vert\langle Q_{it,1},Q_{it,2}\rangle\right\vert>m_1T(\tau_i-t)$ for $1\leq t<\tau_i$ from~\eqref{eq4.2}. Then, by the Cauchy-Schwarz inequality, we can prove that
\[
\left\vert\frac{\langle P_{it,1}, P_{it,2}\rangle}{\langle Q_{it,1}, Q_{it,2}\rangle}\right\vert\leq\frac{\Vert P_{it,1}\Vert\cdot \Vert P_{it,2}\Vert}{m_1T(\tau_i-t)}=\frac{1}{m_1(\tau_i-t)}\frac{\Vert P_{it,1}\Vert}{T^{1/2}}\cdot \frac{\Vert P_{it,2}\Vert}{T^{1/2}}.
\]
Following the proof of Proposition~\ref{prop:A.2} in Appendix~B, we may show that
\[
{\sf P}\left(\max_{i\in{\mathcal C}_\bullet}\max_{1\leq t\leq \tau_i-\theta[\ln(N\vee T)]^{1+\zeta}}\Vert P_{it,1}\Vert>T^{1/2}[\ln(N\vee T)]^{1/2+\zeta/3}\right)\rightarrow0,
\]
and
\[
{\sf P}\left(\max_{i\in{\mathcal C}_\bullet}\max_{1\leq t\leq \tau_i-\theta[\ln(N\vee T)]^{1+\zeta}}\Vert P_{it,2}\Vert>T^{1/2}[\ln(N\vee T)]^{1/2+\zeta/3}\right)\rightarrow0.
\]
Hence, we can prove that
\begin{equation}\label{eqA.14}
\max_{i\in{\mathcal C}_\bullet}\max_{1\leq t\leq \tau_i-\theta[\ln(N\vee T)]^{1+\zeta}}\left\vert\frac{\langle P_{it,1}, P_{it,2}\rangle}{\langle Q_{it,1}, Q_{it,2}\rangle}\right\vert=o_P(1).
\end{equation}
Similarly, we may also show that
\begin{equation}\label{eqA.15}
\max_{i\in{\mathcal C}_\bullet}\max_{1\leq t\leq \tau_i-\theta[\ln(N\vee T)]^{1+\zeta}}\left\vert\frac{\langle P_{it,1}, Q_{it,2}\rangle}{\langle Q_{it,1}, Q_{it,2}\rangle}\right\vert=o_P(1)
\end{equation}
and
\begin{equation}\label{eqA.16}
\max_{i\in{\mathcal C}_\bullet}\max_{1\leq t\leq \tau_i-\theta[\ln(N\vee T)]^{1+\zeta}}\left\vert\frac{\langle Q_{it,1}, P_{it,2}\rangle}{\langle Q_{it,1}, Q_{it,2}\rangle}\right\vert=o_P(1).
\end{equation}
In virtue of~\eqref{eqA.13}--(\ref{eqA.16}), we readily have that
\begin{equation}\label{eqA.17}
{\sf P}\left(\min_{i\in{\mathcal C}_\bullet}\max_{1\leq t\leq \tau_i-\theta[\ln(N\vee T)]^{1+\zeta}}\left( \Vert Z_{it}\Vert^2-\Vert Z_{i\tau_i}\Vert^2\right)>-C\right)\rightarrow0
\end{equation}
for all $C>0$. For the case of $\tau_i<t\leq T$, we can also prove that
\begin{equation}\label{eqA.18}
{\sf P}\left(\min_{i\in{\mathcal C}_\bullet}\max_{ \tau_i+\theta[\ln(N\vee T)]^{1+\zeta}\leq t\leq T}\left( \Vert Z_{it}\Vert^2-\Vert Z_{i\tau_i}\Vert^2\right)>-C\right)\rightarrow0
\end{equation}
in exactly the same way. With~\eqref{eqA.17} and~\eqref{eqA.18}, we prove~\eqref{eq4.5} letting $\theta$ be arbitrarily small.\hfill$\Box$

\medskip

\noindent{\bf Proof of Theorem~\ref{thm:3}}.\ Let $\Lambda_{NT}$ denote the event 
\[
\left\{\widehat{\mathcal C}_\bullet={\mathcal C}_\bullet,\ \ \max_{i\in{\mathcal C}_\bullet}\left\vert\widehat{\tau}_i-\tau_i\right\vert\leq \theta_{NT}\right\},
\]
where $\theta_{NT}=\theta[\ln(N\vee T)]^{1+\zeta}$ with $\theta$ being any positive number. By Assumption~\ref{ass:2}(i)(ii), we may verify the conditions in Theorem~\ref{thm:2} and consequently ${\sf P}(\Lambda_{NT})\rightarrow1$. Therefore, we next prove the theorem conditional on $\Lambda_{NT}$. By the construction~\eqref{eq4.7} and Assumption~\ref{ass:2}(i), we may show that
\[
{\sf P}\left(\widehat{\mathcal C}(k|K_0)={\mathcal C}(b_k),\ k=1,\cdots,K_0\ |\ \Lambda_{NT}\right)\rightarrow1
\]
and $\widehat{\mathcal C}(b_k)=\widehat{\mathcal C}(k|K_0)$ when $\widehat{K}=K_0$. This, together with~\eqref{eq4.11}, proves~\eqref{eq4.12}. 

It remains to prove~\eqref{eq4.11}. By the definition~\eqref{eq4.10}, we only need to show that 
\begin{equation}\label{eqA.19}
{\sf P}\left({\sf IC}(K)>{\sf IC}(K_0),\ 1\leq K\neq K_0\leq \overline{K}\right)\rightarrow1.
\end{equation}
We next consider the two cases: $K>K_0$ and $K<K_0$, separately. 

When $K_0+1\leq K\leq \overline{K}$, conditional on $\Lambda_{NT}$, $\widehat{\tau}_{k|K}$ defined in~\eqref{eq4.8} is a consistent estimate of 
\[\tau_{k|K}=\frac{1}{\left\vert\widehat{\mathcal C}(k|K)\right\vert}\sum_{i\in\widehat{\mathcal C}(k|K)}\tau_i,\]
and there exists $j\in\{1,\cdots,K_0\}$ such that $\widehat{\mathcal C}(k|K)$ is a consistent estimate of ${\mathcal C}(b_j)$ or its subset. Conditional on $\Lambda_{NT}$, we have $\tau_{k|K}-\theta_{NT}\leq\widehat{\tau}_{k|K}\leq \tau_{k|K}+\theta_{NT}$. If $\tau_{k|K}-\theta_{NT}\leq\widehat{\tau}_{k|K}\leq \tau_{k|K}$, we may show that
\begin{eqnarray}
\widehat{\mu}_{i,k|K}&=&\frac{1}{\widehat{\tau}_{k|K}}\left(\sum_{t=1}^{\tau_{k|K}}\mu_i+\sum_{t=1}^{\widehat{\tau}_{k|K}}\epsilon_{it}-\sum_{t=\widehat{\tau}_{k|K}+1}^{\tau_{k|K}}\mu_i\right)\notag\\
&=&\mu_i+\frac{1}{\widehat{\tau}_{k|K}}\sum_{t=1}^{\widehat{\tau}_{k|K}}\epsilon_{it}+O_P\left(\frac{[\ln(N\vee T)]^{1+\zeta}}{T}\right),\label{eqA.20}
\end{eqnarray}
and 
\begin{equation}\label{eqA.21}
\widehat{\delta}_{i,k|K}=\delta_i+\frac{1}{T-\widehat{\tau}_{k|K}}\sum_{t=\widehat{\tau}_{k|K}+1}^T\epsilon_{it}-\frac{1}{\widehat{\tau}_{k|K}}\sum_{t=1}^{\widehat{\tau}_{k|K}}\epsilon_{it}+O_P\left(\frac{[\ln(N\vee T)]^{1+\zeta}}{T}\right).
\end{equation}
Similarly,~\eqref{eqA.20} and~\eqref{eqA.21} also hold when $\tau_{k|K}+1\leq\widehat{\tau}_{k|K}\leq \tau_{k|K}+\theta_{NT}$. With~\eqref{eqA.20} and~\eqref{eqA.21}, for $\widehat\nu_{it,k|K}$ defined in~\eqref{eq4.8}, we have
\begin{equation}\label{eqA.22}
\widehat{\nu}_{it,k|K}=\left\{
\begin{array}{ll}
\mu_i+\frac{1}{\widehat{\tau}_{k|K}}\sum_{t=1}^{\widehat{\tau}_{k|K}}\epsilon_{it}+O_P\left(\frac{[\ln(N\vee T)]^{1+\zeta}}{T}\right),&t\leq\widehat{\tau}_{k|K},\\
\mu_i+\delta_i+\frac{1}{T-\widehat{\tau}_{k|K}}\sum_{t=\widehat{\tau}_{k|K}+1}^T\epsilon_{it}+O_P\left(\frac{[\ln(N\vee T)]^{1+\zeta}}{T}\right),&t>\widehat{\tau}_{k|K}.
\end{array}
\right.
\end{equation}
By~\eqref{eq4.6},~\eqref{eqA.22} and Proposition~\ref{prop:A.2}, for $i\in\widehat{\mathcal C}(k|K)$, we can prove that
\begin{eqnarray}
&&\frac{1}{T}\sum_{i\in\widehat{\mathcal C}(k|K)}\sum_{t=1}^T\left\Vert X_{it}-\widehat{\nu}_{it,k|K}\right\Vert^2\notag\\
&=&\frac{1}{T}\sum_{i\in\widehat{\mathcal C}(k|K)}\left(\sum_{t=1}^{\lfloor \tau_{k|K}-\theta_{NT}\rfloor}+\sum_{t=\lfloor \tau_{k|K}-\theta_{NT}\rfloor+1}^{\lfloor \tau_{k|K}+\theta_{NT}\rfloor}+\sum_{t=\lfloor \tau_{k|K}+\theta_{NT}\rfloor+1}^T\right)\left\Vert X_{it}-\widehat{\nu}_{it,k|K}\right\Vert^2\notag\\
&=&\frac{1}{T}\sum_{i\in\widehat{\mathcal C}(k|K)}\sum_{t=1}^{\lfloor \tau_{k|K}-\theta_{NT}\rfloor}\left\Vert \epsilon_{it}-\frac{1}{\widehat{\tau}_{k|K}}\sum_{t=1}^{\widehat{\tau}_{k|K}}\epsilon_{it}\right\Vert^2+\frac{1}{T}\sum_{i\in\widehat{\mathcal C}(k|K)}\sum_{t=\lfloor \tau_{k|K}+\theta_{NT}\rfloor+1}^T\left\Vert \epsilon_{it}-\frac{1}{T-\widehat{\tau}_{k|K}}\sum_{t=\widehat{\tau}_{k|K}+1}^T\epsilon_{it}\right\Vert^2\notag\\
&&+O_P\left(\big|\widehat{\mathcal C}(k|K)\big|\cdot\frac{[\ln(N\vee T)]^{1+\zeta}}{T}\right)\notag\\
&=&\frac{1}{T}\sum_{i\in\widehat{\mathcal C}(k|K)}\sum_{t=1}^{\lfloor \tau_{k|K}-\theta_{NT}\rfloor}\left\Vert \epsilon_{it}\right\Vert^2+\frac{1}{T}\sum_{i\in\widehat{\mathcal C}(k|K)}\sum_{t=\lfloor \tau_{k|K}+\theta_{NT}\rfloor+1}^T\left\Vert \epsilon_{it}\right\Vert^2+O_P\left(\big|\widehat{\mathcal C}(k|K)\big|\cdot\frac{[\ln(N\vee T)]^{1+\zeta}}{T}\right)\notag\\
&=&\frac{1}{T}\sum_{i\in\widehat{\mathcal C}(k|K)}\sum_{t=1}^T\left\Vert \epsilon_{it}\right\Vert^2+O_P\left(\big|\widehat{\mathcal C}(k|K)\big|\cdot\frac{[\ln(N\vee T)]^{1+\zeta}}{T}\right),\notag
\end{eqnarray}
which, together with the fact $\big|\widehat{\mathcal C}_\bullet\big|=\sum_{k=1}^{K}\big|\widehat{\mathcal C}(k|K)\big|$, indicates that
\begin{eqnarray}
V(K) &=& \frac{1}{T\big|\widehat{\mathcal C}_\bullet\big|}\sum_{k=1}^K\sum_{i\in\widehat{\mathcal C}(k|K)}\sum_{t=1}^T\left\Vert \epsilon_{it}\right\Vert^2+O_P\left(\frac{[\ln(N\vee T)]^{1+\zeta}}{T}\right)\notag\\
&=& V(K_0)+O_P\left(\frac{[\ln(N\vee T)]^{1+\zeta}}{T}\right).\label{eqA.23}
\end{eqnarray}
By~\eqref{eqA.23} and Assumption~\ref{ass:2}(iii), we readily have
\begin{equation}\label{eqA.24}
{\sf P}\left({\sf IC}(K)>{\sf IC}(K_0),\ K_0+1\leq K\leq \overline{K}\ |\ \Lambda_{NT}\right)\rightarrow1.
\end{equation}

Let $\Lambda_{NT}^\ast=\left\{\Lambda_{NT},\ \widehat{\mathcal C}(k|K_0)={\mathcal C}(b_k),\ k=1,\cdots,K_0\right\}$. It is easy to verify that ${\sf P}\left(\Lambda_{NT}^\ast\right)\rightarrow1$ by~\eqref{eq4.12}. When $1\leq K\leq K_0-1$, conditional on $\Lambda_{NT}^\ast$, at least two clusters would be falsely merged. Without loss of generality, we consider $K=K_0-1$ and assume that, conditional on $\Lambda_{NT}^\ast$, ${\mathcal C}(b_1)$ and ${\mathcal C}(b_2)$ are falsely merged, i.e., $\widehat{\mathcal C}(1|K_0-1)={\mathcal C}(b_1)\cup{\mathcal C}(b_2)$. In this case, $\widehat{\tau}_{1|K_0-1}$ is a biased change point estimate but can be seen as an estimate of a weighted average of $b_1$ and $b_2$ defined as
\[
\tau_{1|K_0-1}=\frac{\vert{\mathcal C}(b_1)\vert}{\vert{\mathcal C}(b_1)\vert+\vert{\mathcal C}(b_2)\vert}b_1+\frac{\vert{\mathcal C}(b_2)\vert}{\vert{\mathcal C}(b_1)\vert+\vert{\mathcal C}(b_2)\vert}b_2.
\]
By Assumption~\ref{ass:2}(i), we may write $b_1<\tau_{1|K_0-1}=c_\ast T<b_2$ with $c_1<c_\ast<c_2$. For $i\in{\mathcal C}(b_1)$, we may show that
\begin{eqnarray}
\widehat{\mu}_{i,1|K_0-1}&=&\frac{1}{\widehat{\tau}_{1|K_0-1}}\left[\sum_{t=1}^{b_1}\mu_i+\sum_{t=b_1+1}^{\widehat{\tau}_{1|K_0-1}}(\mu_i+\delta_i)+\sum_{t=1}^{\widehat{\tau}_{1|K_0-1}}\epsilon_{it}\right]\notag\\
&=&\frac{c_1}{c_\ast}\mu_i+\frac{c_\ast-c_1}{c_\ast}(\mu_i+\delta_i)+o_P(1),\notag
\end{eqnarray}
and 
\begin{eqnarray}
\widehat{\delta}_{i,1|K_0-1}&=&\frac{1}{T-\widehat{\tau}_{1|K_0-1}}\left[\sum_{t=\widehat{\tau}_{1|K_0-1}+1}^T(\mu_i+\delta_i)+\sum_{t=\widehat{\tau}_{1|K_0-1}+1}^T\epsilon_{it}\right]-\widehat{\mu}_{i,1|K_0-1}\notag\\
&=&\mu_i+\delta_i-\frac{c_1}{c_\ast}\mu_i-\frac{c_\ast-c_1}{c_\ast}(\mu_i+\delta_i)+o_P(1)\notag\\
&=&\frac{c_1}{c_\ast}\delta_i+o_P(1).\notag
\end{eqnarray}
Consequently, for $i\in{\mathcal C}(b_1)$, we can prove that 
\begin{eqnarray}
&&\frac{1}{T}\sum_{t=1}^T\left\Vert X_{it}-\widehat{\nu}_{it,1|K_0-1}\right\Vert^2\notag\\
&=&\frac{1}{T}\left(\sum_{t=1}^{b_1}+\sum_{t=b_1+1}^{\widehat{\tau}_{1|K_0-1}}+\sum_{t=\widehat{\tau}_{1|K_0-1}+1}^T\right)\left\Vert X_{it}-\widehat{\nu}_{it,1|K_0-1}\right\Vert^2\notag\\
&=&\frac{1}{T}\sum_{t=1}^{b_1}\left\Vert \epsilon_{it}-\frac{c_\ast-c_1}{c_\ast}\delta_i \right\Vert^2+\sum_{t=b_1+1}^{\widehat{\tau}_{1|K_0-1}}\left\Vert \epsilon_{it}+\frac{c_1}{c_\ast}\delta_i\right\Vert^2+\sum_{t=\widehat{\tau}_{1|K_0-1}+1}^T\left\Vert \epsilon_{it}\right\Vert^2+o_P(1)\notag\\
&=&\frac{1}{T}\sum_{t=1}^{T}\left\Vert \epsilon_{it}\right\Vert^2+\gamma_i+o_P(1),\label{eqA.25}
\end{eqnarray}
where $\gamma_i=\frac{c_1(c_\ast-c_1)}{c_\ast}\Vert\delta_i\Vert^2>0$. Analogously, for $i\in{\mathcal C}(b_2)$, there also exits $\gamma_i>0$ such that
\begin{equation}\label{eqA.26}
\frac{1}{T}\sum_{t=1}^T\left\Vert X_{it}-\widehat{\nu}_{i,1|K_0-1,t}\right\Vert^2=\frac{1}{T}\sum_{t=1}^{T}\left\Vert \epsilon_{it}\right\Vert^2+\gamma_i+o_P(1).
\end{equation}
Then, by~\eqref{eqA.25} and~\eqref{eqA.26}, we have
\[
{\sf P}\left({\sf IC}(K_0-1)-{\sf IC}(K_0)>0\ |\ \Lambda_{NT}^\ast\right)\rightarrow1.
\]
More generally (but with similar arguments), we can prove that
\begin{equation}\label{eqA.27}
{\sf P}\left({\sf IC}(K)>{\sf IC}(K_0),\ 1\leq K\leq K_0-1\ |\ \Lambda_{NT}^\ast\right)\rightarrow1.
\end{equation}
By~\eqref{eqA.24} and~\eqref{eqA.27}, and noting that ${\sf P}(\Lambda_{NT})\rightarrow1$ and ${\sf P}(\Lambda_{NT}^\ast)\rightarrow1$, we can prove~\eqref{eqA.19}.\hfill$\Box$

\medskip

\noindent{\bf Proof of Theorem~\ref{thm:4}}.\ \ By~\eqref{eq4.12} in Theorem~\ref{thm:3}, we only need to prove~\eqref{eq4.15} conditional on that $\widehat{\mathcal C}(b_k)={\mathcal C}(b_k)$. For each $k=1,\cdots,K_0$ and any $t=1,\cdots, T-1$, we observe that 
\begin{eqnarray}
&&Z(t; {\mathcal C}(b_k))-Z(b_k; {\mathcal C}(b_k))\notag\\
&=&\left\{Z(t; {\mathcal C}(b_k))-{\sf E}[ Z(t; {\mathcal C}(b_k))]\right\}-\left\{Z(b_k; {\mathcal C}(b_k))-{\sf E}[ Z(b_k; {\mathcal C}(b_k))]\right\}\notag\\
&&+\left\{{\sf E}[ Z(t; {\mathcal C}(b_k))]-{\sf E}[ Z(b_k; {\mathcal C}(b_k))]\right\}\notag\\
&\leq& 2\max_{1\leq t\leq T-1}\left\vert Z(t; {\mathcal C}(b_k))-{\sf E}[ Z(t; {\mathcal C}(b_k))]\right\vert +\left\{{\sf E}[ Z(t; {\mathcal C}(b_k))]-{\sf E}[ Z(b_k; {\mathcal C}(b_k))]\right\},\label{eqA.28}
\end{eqnarray}
where $Z(t; {\mathcal C}(b_k))$ is defined in Proposition~\ref{prop:A.3}. By the definition of $\widehat{b}_k$ in~\eqref{eq4.13} and Propositions~\ref{prop:A.3} and~\ref{prop:A.4}, we readily have that 
\begin{equation}\label{eqA.29}
m_0|\widehat{b}_k-b_k|\sum_{i\in{\mathcal C}(b_k)}\Vert\delta_i\Vert^2=O_P\left(|{\mathcal C}(b_k)|^{1/2}+T^{1/2}\left(\sum_{i\in{\mathcal C}(b_k)}\Vert\delta_i\Vert^2\right)^{1/2}\right)
\end{equation}
under $H_A$, where $m_0$ is defined in Proposition~\ref{prop:A.4}. Letting 
\[\delta_{{\mathcal C}(b_k)}=\left(\frac{1}{|{\mathcal C}(b_k)|^{1/2}}\sum_{i\in{\mathcal C}(b_k)}\Vert\delta_i\Vert^2\right)^{-1}\rightarrow0,\]
by~\eqref{eqA.29}, we have
\begin{equation}\label{eqA.30}
\left|\widehat{b}_k-b_k\right|=O_P\left(\delta_{{\mathcal C}(b_k)}+T^{1/2}|{\mathcal C}( b_k)|^{-1/4}\delta_{{\mathcal C}(b_k)}^{1/2}\right)=o_P\left(\varpi_k\right),\ \ k=1,\cdots,K_0
\end{equation}
where $\varpi_k=T^{1/2}|{\mathcal C}(b_k)|^{-1/4}$. Let ${\cal N}_k(\epsilon)$ be defined in Proposition~\ref{prop:A.5}, from~\eqref{eqA.30}, we have
\begin{eqnarray}
{\sf P}\left(\widehat{b}_k\neq b_k\right)&\leq&{\sf P}\left(\widehat{b}_k\in{\cal N}_k(\epsilon),\ \widehat{b}_k\neq b_k\right)+{\sf P}\left(\widehat{b}_k\not\in{\cal N}_k(\epsilon)\right)\notag\\
&=&{\sf P}\left(\widehat{b}_k\in{\cal N}_k(\epsilon),\ \widehat{b}_k\neq b_k\right)+o(1)\label{eqA.31}
\end{eqnarray}
for any $k=1,\cdots,K_0$. Furthermore, Proposition~\ref{prop:A.5} indicates that 
\begin{equation}\label{eqA.32}
{\sf P}\left(\widehat{b}_k\in{\cal N}_k(\epsilon),\ \widehat{b}_k\neq b_k\right)\rightarrow0,\ \ k=1,\cdots,K_0.
\end{equation}
In view of~\eqref{eqA.31} and~\eqref{eqA.32}, we complete the proof of Theorem~\ref{thm:4}. \hfill $\Box$

\bigskip


\newpage

\begin{center}

{\Large Supplement to ``Detection and Estimation of Structural Breaks in High-Dimensional Functional Time Series"}

\end{center}

\bigskip


\appendix

\section*{\bf\Large Appendix B:\ \ Proofs of technical results}\label{app:B}
\renewcommand{\theequation}{B.\arabic{equation}}
\setcounter{equation}{0}

In this appendix, we give the detailed proofs of the technical results which play a crucial role in proving the main asymptotic theorems in Appendix A. Let $M$ be a generic positive constant whose value is allowed to change from line to line.

\medskip

\noindent{\bf Proof of Proposition A.1}.\ \ Note that
\begin{equation}\label{eqB.1}
\widetilde{S}_{NT}^\epsilon(x;u)=N^{1/2}\sum_{s=1}^{\lfloor Tx\rfloor}\widetilde{\epsilon}_{s}(u)=\frac{1}{\sqrt{N}}\sum_{s=1}^{\lfloor Tx\rfloor}\sum_{i=1}^N \epsilon_{is}(u)
\end{equation}
using the definition of $\widetilde{\epsilon}_{s}(u)$. By the Beveridge-Nelson decomposition \citep[e.g.,][]{PS92}, we may show that
\begin{equation}\label{eqB.2}
\sum_{s=1}^t \epsilon_{is}={\mathbf A}_i\sum_{s=1}^t \eta_{is}+\widetilde{\epsilon}_{i0}-\widetilde{\epsilon}_{it}
\end{equation}
for any positive integer $t$, where $\epsilon_{is}=(\epsilon_{is}(u): u\in{\mathbb C})$, $\eta_{is}=(\eta_{is}(u): u\in{\mathbb C})$, ${\mathbf A}_i=\sum_{j=0}^\infty {\mathbf A}_{ij}$ and $\widetilde{\epsilon}_{it}=\sum_{j=0}^\infty \widetilde{\mathbf A}_{ij}\eta_{i,t-j}$ with $\widetilde{\mathbf A}_{ij}=\sum_{k=j+1}^\infty {\mathbf A}_{ik}$. Combining (\ref{eqB.1}) and (\ref{eqB.2}), we have
\begin{equation}\label{eqB.3}
\widetilde{S}_{NT}^\epsilon(x;u)=\sum_{s=1}^{\lfloor Tx\rfloor}\frac{1}{\sqrt{N}}\sum_{i=1}^N \epsilon_{is}(u)=\sum_{s=1}^{\lfloor Tx\rfloor}\frac{1}{\sqrt{N}}\sum_{i=1}^N {\mathbf A}_i\eta_{is}(u)+\frac{1}{\sqrt{N}}\sum_{i=1}^N\left[\widetilde{\epsilon}_{i0}(u)-\widetilde{\epsilon}_{i\lfloor Tx\rfloor}(u)\right].
\end{equation}

Let $\widetilde{\eta}_{s}=\frac{1}{\sqrt{N}}\sum_{i=1}^N {\mathbf A}_i\eta_{is}$ which is i.i.d. over $s$ with mean zero and a positive definite covariance operator by Assumption 1. Using the weak invariance principle for independent partial sums of random functions \citep[e.g.,][]{dA82, DP83, BHR13}, we may show that 
\begin{equation}\label{eqB.4}
\sup_{0\leq x\leq 1}\int _{\mathbb C} \left[\frac{1}{T^{1/2}}\sum_{s=1}^{\lfloor Tx\rfloor}\frac{1}{\sqrt{N}}\sum_{i=1}^N {\mathbf A}_i\eta_{is}(u)-G_{NT}(x;u)\right]^2du=o_P(1).
\end{equation}
With (\ref{eqB.4}), to complete the proof of (A.1), we only need to prove that 
\begin{equation}\label{eqB.5}
\max_{1\leq t\leq T}\frac{1}{\sqrt{N}}\left\Vert\sum_{i=1}^N\widetilde{\epsilon}_{it}\right\Vert=o_P\left(T^{1/2}\right).
\end{equation}
In fact, by the Burkholder-Rosenthal inequality for the random elements in the Hilbert space \citep[e.g.,][]{dA81, Os12}, (2.4) and the moment condition (2.6), we have
\begin{equation}\label{eqB.6}
\frac{1}{N^{1+\iota/2}}{\sf E}\left[\left\Vert\sum_{i=1}^N\widetilde{\epsilon}_{it}\right\Vert^{2+\iota}\right]\leq M,
\end{equation}
which, together with the Bonferroni and Markov inequalities, leads to
\[
{\sf P}\left(\max_{1\leq t\leq T}\frac{1}{\sqrt{N}}\left\Vert\sum_{i=1}^N\widetilde{\epsilon}_{it}\right\Vert>\varepsilon T^{1/2}\right)\leq \frac{\varepsilon^{-(2+\iota)}}{(NT)^{1+\iota/2}}\sum_{t=1}^T {\sf E}\left[\left\Vert\sum_{i=1}^N\widetilde{\epsilon}_{it}\right\Vert^{2+\iota}\right]=o\left(T^{-\iota/2}\right)=o(1)
\]
for any $\varepsilon>0$. This proves (\ref{eqB.5}), thus completing the proof of Proposition A.1.\hfill$\Box$

\medskip

\noindent{\bf Proof of Proposition A.2}.\ \ By (\ref{eqB.2}), we may show that, under $H_0$,
\begin{eqnarray}
Z_{iT}(x;u)&=&\frac{1}{\sqrt{T}}\left[\left(1-\frac{\lfloor Tx\rfloor}{T}\right)\widetilde{\epsilon}_{i0}(u)-\left(1-\frac{\lfloor Tx\rfloor}{T}\right)\widetilde{\epsilon}_{i\lfloor Tx\rfloor}(u)\right]\notag\\
&&+\frac{1}{\sqrt{T}}\left[{\mathbf A}_i\sum_{s=1}^{\lfloor Tx\rfloor}\eta_{is}(u)-\frac{\lfloor Tx\rfloor}{T}{\mathbf A}_i\sum_{s=1}^T\eta_{is}(u)\right].\notag
\end{eqnarray}
Then we readily have that
\begin{equation}\label{eqB.7}
\sup_{0\leq x\leq 1}\int_{\mathbb C} Z_{iT}^2(x;u)du\leq M\left(\frac{1}{T}\max_{1\leq t\leq T}\left\Vert\sum_{s=1}^t\eta_{is}\right\Vert^2+\frac{1}{T}\max_{1\leq t\leq T}\left\Vert\widetilde{\epsilon}_{it}\right\Vert^2\right).
\end{equation}
Hence, in order to prove (A.2), it is sufficient to show
\begin{equation}\label{eqB.8}
{\sf P}\left(\max_{1\leq i\leq N}\max_{1\leq t\leq T} \left\Vert\sum_{s=1}^t\eta_{is}\right\Vert^2>T\cdot\xi_{NT}\right)\rightarrow0
\end{equation}
and
\begin{equation}\label{eqB.9}
{\sf P}\left(\max_{1\leq i\leq N}\max_{1\leq t\leq T} \left\Vert\widetilde{\epsilon}_{it}\right\Vert^2>T\cdot\xi_{NT}\right)\rightarrow0.
\end{equation}

We first prove (\ref{eqB.9}) and then (\ref{eqB.8}). Applying the Burkholder-Rosenthal inequality for the random elements in the Hilbert space as in (\ref{eqB.6}), we may show that ${\sf E}\left\Vert\widetilde{\epsilon}_{it}\right\Vert^{2(\kappa+1)}\leq M$ for $\kappa\geq0$ defined as in Theorem 1. Then, using the Bonferroni and Markov inequalities, we can prove that
\begin{eqnarray}
&&{\sf P}\left(\max_{1\leq i\leq N}\max_{1\leq t\leq T} \left\Vert\widetilde{\epsilon}_{it}\right\Vert^2>T\cdot\xi_{NT}\right)\notag\\
&\leq&\sum_{i=1}^N\sum_{t=1}^T{\sf P}\left( \left\Vert\widetilde{\epsilon}_{it}\right\Vert^2>T\cdot\xi_{NT}\right)\leq\sum_{i=1}^N\sum_{t=1}^T\frac{{\sf E}\left\Vert\widetilde{\epsilon}_{it}\right\Vert^{2(\kappa+1)}}{\left(T\cdot\xi_{NT}\right)^{\kappa+1}}\notag\\
&\leq&M\left(NT^{-\kappa}\right)\xi_{NT}^{-(\kappa+1)}\rightarrow0,\notag
\end{eqnarray}
where the facts: $N=O\left(T^\kappa\right)$ and $\xi_{NT}\rightarrow\infty$, have been used.

Let
\[\overline{\eta}_{is}=\eta_{is}I\left(\Vert\eta_{is}\Vert\leq\xi_{NT}^{1/2}\right),\ \ 
\widehat{\eta}_{is}=\eta_{is}I\left(\Vert\eta_{is}\Vert>\xi_{NT}^{1/2}\right).
\]
To prove (\ref{eqB.8}), we only need to show 
\begin{equation}\label{eqB.10}
{\sf P}\left(\max_{1\leq i\leq N}\max_{1\leq t\leq T} \left\Vert\sum_{s=1}^t\left(\overline\eta_{is}-{\sf E}\overline{\eta}_{is}\right)\right\Vert^2>\frac{1}{4}T\cdot\xi_{NT}\right)\rightarrow0,
\end{equation}
and
\begin{equation}\label{eqB.11}
{\sf P}\left(\max_{1\leq i\leq N}\max_{1\leq t\leq T} \left\Vert\sum_{s=1}^t\left(\widehat\eta_{is}-{\sf E}\widehat{\eta}_{is}\right)\right\Vert^2>\frac{1}{4}T\cdot\xi_{NT}\right)\rightarrow0.
\end{equation}

As $\max_{1\leq i\leq N}{\sf E}\left[\exp\left\{c_\eta\Vert\eta_{it}\Vert^2\right\}\right]<\infty$ in Assumption 1(ii), we may prove that
\begin{eqnarray}
{\sf E}\left\Vert\widehat{\eta}_{is}\right\Vert&\leq&\left({\sf E}\left\Vert {\eta}_{is}\right\Vert^2\right)^{1/2}\left[{\sf P}\left(\Vert\eta_{is}\Vert>\xi_{NT}^{1/2}\right)\right]^{1/2}\notag\\
&\leq&M\exp\left\{-\frac{1}{2}\xi_{NT}\right\}=O\left((N\vee T)^{-\frac{1}{2}c_\xi\ln\ln(N\vee T)}\right)\notag\\
&=&o\left(\left[T^{-1}\xi_{NT}\right]^{1/2}\right),\notag
\end{eqnarray}
which indicates that
\begin{eqnarray}
&&{\sf P}\left(\max_{1\leq i\leq N}\max_{1\leq t\leq T} \left\Vert\sum_{s=1}^t\left(\widehat\eta_{is}-{\sf E}\widehat{\eta}_{is}\right)\right\Vert^2>\frac{1}{4}T\cdot\xi_{NT}\right)\notag\\
&\leq&{\sf P}\left(\max_{1\leq i\leq N}\max_{1\leq t\leq T} \left\Vert\sum_{s=1}^t\widehat\eta_{is}\right\Vert^2>\frac{1}{5}T\cdot\xi_{NT}\right)\notag\\
&\leq&\sum_{i=1}^N\sum_{t=1}^T{\sf P}\left(\left\Vert \eta_{it}\right\Vert>\xi_{NT}^{1/2}\right)\leq M\sum_{i=1}^N\sum_{t=1}^T \frac{{\sf E}\exp\left\{c_\eta\cdot\Vert\eta_{it}\Vert^2\right\}}{\exp\left\{c_\eta\cdot \xi_{NT}\right\}}\notag\\
&=&O\left((N\vee T)^{2-c_\eta c_\xi\ln\ln(N\vee T)}\right)=o(1),\notag
\end{eqnarray}
proving (\ref{eqB.11}).

On the other hand, by the Bernstein-type inequality for independent random elements in the Hilbert space \citep[e.g., Theorem 2.6(2) in][]{B00}, we may show that
\begin{eqnarray}
&&{\sf P}\left(\max_{1\leq i\leq N}\max_{1\leq t\leq T} \left\Vert\sum_{s=1}^t\left(\overline\eta_{is}-{\sf E}\overline{\eta}_{is}\right)\right\Vert^2>\frac{1}{4}T\cdot\xi_{NT}\right)\notag\\
&\leq&\sum_{i=1}^N\sum_{t=1}^T{\sf P}\left(\left\Vert\sum_{s=1}^t\left(\overline\eta_{is}-{\sf E}\overline{\eta}_{is}\right)\right\Vert^2>\frac{1}{4}T\cdot\xi_{NT}\right)\notag\\
&\leq&M(NT)\cdot\exp\left\{-m_\diamond\ln(N\vee T)\ln\ln(N\vee T)\right\}\notag\\
&=&O\left((N\vee T)^{2-m_\diamond\ln\ln(N\vee T)}\right)=o(1),\notag
\end{eqnarray}
where $m_\diamond$ is a positive constant. This proves (\ref{eqB.10}). The proof of (\ref{eqB.8}) is completed.\hfill$\Box$

\medskip

\noindent{\bf Proof of Proposition A.3}.\ \ Under $H_A$, we note that
\begin{equation}\label{eqB.12}
Z(t; {\cal C}(b_k))=\sum_{i\in{\cal C}(b_k)}\left\{\int_{\mathbb C} \left[Z_{iT}^\epsilon(t/T;u)\right]^2du+D_k^2(t) \int_{\mathbb C}\delta_i^2(u)du-2D_k(t)\int_{\mathbb C} Z_{iT}^\epsilon(t/T;u)\delta_i(u)du\right\},
\end{equation}
where $Z_{iT}^\epsilon(t/T;u)$ is defined in the proof of Theorem 1(ii) and
\[D_k(t)=\left\{
\begin{array}{cc}
\sqrt{T}\left(\frac{t}{T}\frac{T-b_k}{T}\right),&t\leq b_k,\\
\sqrt{T}\left(\frac{b_k}{T}\frac{T-t}{T}\right),&t> b_k.
\end{array}\right.
\]
As the second term on RHS of (\ref{eqB.12}) is non-random and the mean of the third term is zero by Assumption 1, we then have
\begin{eqnarray}
Z(t; {\cal C}(b_k))-{\sf E}[Z(t; {\cal C}(b_k))]&=&\sum_{i\in{\cal C}(b_k)}\left(\int_{\mathbb C}\left\{ \left[Z_{iT}^\epsilon(t/T;u)\right]^2-{\sf E}\left[Z_{iT}^\epsilon(t/T;u)\right]^2\right\}du\right.\notag\\
&&\left.-2D_k(t)\int_{\mathbb C} Z_{iT}^\epsilon(t/T;u)\delta_i(u)du\right).\label{eqB.13}
\end{eqnarray}

As $Z_{iT}^\epsilon(t/T;u)$ is independent over $i$, following the proof of Theorem 1 in \cite{HH12} (with some modifications) and using the continuous mapping theorem \citep[e.g.,][]{Bi68}, we may show that 
\begin{equation}\label{eqB.14}
\max_{1\leq t\leq T}\left\vert \sum_{i\in{\cal C}(b_k)}\int_{\mathbb C}\left\{ \left[Z_{iT}^\epsilon(t/T;u)\right]^2-{\sf E}\left[Z_{iT}^\epsilon(t/T;u)\right]^2\right\}du\right\vert=O_P\left(|{\cal C}(b_k)|^{1/2}\right)
\end{equation}
and
\begin{equation}\label{eqB.15}
\max_{1\leq t\leq T}\left\vert \sum_{i\in{\cal C}(b_k)}\int_{\mathbb C} Z_{iT}^\epsilon(t/T;u)\delta_i(u)du\right\vert=O_P\left(\left(\sum_{i\in{\cal C}(b_k)} \Vert\delta_i\Vert^2\right)^{1/2}\right).
\end{equation}
By Assumption 2(i), we have $D_k(t)=O\left(T^{1/2}\right)$ uniformly over $1\leq t\leq T$, which, together with (\ref{eqB.15}), leads to
\begin{eqnarray}
&&\max_{1\leq t\leq T}\left\vert D_k(t)\sum_{i\in{\cal C}(b_k)}\int_{\mathbb C} Z_{iT}^\epsilon(t/T;u)\delta_i(u)du\right\vert\notag\\
&\leq& \left[\max_{1\leq t\leq T}D_k(t)\right]\max_{0\leq t\leq T}\left\vert\sum_{i\in{\cal C}(b_k)}\int_{\mathbb C} Z_{iT}^\epsilon(t/T;u)\delta_i(u)du\right\vert\notag\\
&=&O_P\left(T^{1/2}\left(\sum_{i\in{\cal C}(b_k)} \Vert\delta_i\Vert^2\right)^{1/2}\right).\label{eqB.16}
\end{eqnarray}
By (\ref{eqB.13}), (\ref{eqB.14}) and (\ref{eqB.16}) and the triangle inequality, we complete the proof of Proposition A.3.\hfill$\Box$

\medskip

\noindent{\bf Proof of Proposition A.4}.\ \ By (\ref{eqB.12}), we readily have that
\begin{eqnarray}
{\sf E}[Z(b_k; {\cal C}(b_k))]-{\sf E}[Z(t; {\cal C}(b_k))]&=&\sum_{i\in{\cal C}(b_k)}\int_{\mathbb C}\left\{ {\sf E}\left[Z_{iT}^\epsilon(b_k/T;u)\right]^2-{\sf E}\left[Z_{iT}^\epsilon(t/T;u)\right]^2\right\}du\notag\\
&&+\left[D_k^2(b_k)-D_k^2(t)\right]\sum_{i\in{\cal C}(b_k)} \Vert\delta_i\Vert^2.\label{eqB.17}
\end{eqnarray}
By Assumption 2(i) and the definition of $D_k(t)$ in the proof of Proposition A.3, there exists a positive constant $m_0$ such that
\[D_k^2(b_k)-D_k^2(t)\geq 2m_0|t-b_k|\ \ \ {\rm for}\ \ t\neq b_k,\ \ k=1,\cdots,K_0.\]
This indicates that 
\begin{equation}\label{eqB.18}
\left[D_k^2(b_k)-D_k^2(t)\right]\sum_{i\in{\cal C}(b_k)} \Vert\delta_i\Vert^2\geq 2m_0|t-b_k| \sum_{i\in{\cal C}(b_k)} \Vert\delta_i\Vert^2.
\end{equation}
On the other hand, by some elementary calculations, we have
\begin{equation}\label{eqB.19}
\sum_{i\in{\cal C}(b_k)}\int_{\mathbb C}\left\{ {\sf E}\left[Z_{iT}^\epsilon(b_k/T;u)\right]^2-{\sf E}\left[Z_{iT}^\epsilon(t/T;u)\right]^2\right\}du=O\left(\frac{|{\cal C}(b_k)|}{T}|t-b_k|\right)
\end{equation}
for any $1\leq t\leq T$. Using the conditions $|{\cal C}_\bullet |=O(T^2)$ and (4.14), we may show that 
\begin{equation}\label{eqB.20}
\frac{|{\cal C}(b_k)|}{T}=o\left(\sum_{i\in{\cal C}(b_k)} \Vert\delta_i\Vert^2\right),
\end{equation}
which, together with (\ref{eqB.17})--(\ref{eqB.19}), completes the proof of Proposition A.4.\hfill$\Box$

\medskip

\noindent{\bf Proof of Proposition A.5}.\ \ By (\ref{eqB.12}) and (\ref{eqB.17}), under $H_A$, we have
\begin{eqnarray}
&&\frac{1}{|{\cal C}(b_k)|^{1/2}}\left[Z(b_k; {\cal C}(b_k))-Z(t; {\cal C}(b_k))\right]\notag\\
&=&\frac{1}{|{\cal C}(b_k)|^{1/2}}\sum_{i\in{\cal C}(b_k)}\int_{\mathbb C}\left\{ \left[Z_{iT}^\epsilon(b_k/T;u)\right]^2-{\sf E}\left[Z_{iT}^\epsilon(b_k/T;u)\right]^2\right\}du\notag\\
&&-\frac{1}{|{\cal C}(b_k)|^{1/2}}\sum_{i\in{\cal C}(b_k)}\int_{\mathbb C}\left\{ \left[Z_{iT}^\epsilon(t/T;u)\right]^2-{\sf E}\left[Z_{iT}^\epsilon(t/T;u)\right]^2\right\}du\notag\\
&&+\frac{1}{|{\cal C}(b_k)|^{1/2}}\sum_{i\in{\cal C}(b_k)}\int_{\mathbb C}\left\{ {\sf E}\left[Z_{iT}^\epsilon(b_k/T;u)\right]^2-{\sf E}\left[Z_{iT}^\epsilon(t/T;u)\right]^2\right\}du\notag\\
&&-\frac{2}{|{\cal C}(b_k)|^{1/2}}\left[D_k(t)\sum_{i\in{\cal C}(b_k)}\int_{\mathbb C}Z_{iT}^\epsilon(t/T;u)\delta_i(u)du-D_k(b_k)\sum_{i\in{\cal C}(b_k)}\int_{\mathbb C}Z_{iT}^\epsilon(b_k/T;u)\delta_i(u)du\right]\notag\\
&&+\frac{1}{|{\cal C}(b_k)|^{1/2}}\left[D_k^2(b_k)-D_k^2(t)\right]\sum_{i\in{\cal C}(b_k)} \Vert\delta_i\Vert^2\label{eqB.21}
\end{eqnarray}
for any $b_k-\varepsilon\varpi_k\leq t< b_k$ and $k=1,\cdots,K_0$. 

By (\ref{eqB.18}) and (4.14), uniformly over $k=1,\cdots,K_0$,
\begin{equation}\label{eqB.22}
\frac{1}{|{\cal C}(b_k)|^{1/2}}\left[D_k^2(b_k)-D_k^2(t)\right]\sum_{i\in{\cal C}(b_k)} \Vert\delta_i\Vert^2>2m_0(b_k-t)\frac{1}{|{\cal C}(b_k)|^{1/2}}\sum_{i\in{\cal C}(b_k)}\Vert\delta_i\Vert^2\rightarrow\infty
\end{equation}
for any $b_k-\varepsilon\varpi_k\leq t< b_k$. We next consider the fourth term on RHS of (\ref{eqB.21}). Note that
\begin{eqnarray}
&&D_k(t)\sum_{i\in{\cal C}(b_k)}\int_{\mathbb C}Z_{iT}^\epsilon(t/T;u)\delta_i(u)du-D_k(b_k)\sum_{i\in{\cal C}(b_k)}\int_{\mathbb C}Z_{iT}^\epsilon(b_k/T;u)\delta_i(u)du\notag\\
&=&D_k(t)\sum_{i\in{\cal C}(b_k)}\int_{\mathbb C}\left[Z_{iT}^\epsilon(t/T;u)-Z_{iT}^\epsilon(b_k/T;u)\right]\delta_i(u)du\notag\\
&&+\left[D_k(t)-D_k(b_k)\right]\sum_{i\in{\cal C}(b_k)}\int_{\mathbb C}Z_{iT}^\epsilon(b_k/T;u)\delta_i(u)du.\notag
\end{eqnarray}
Similarly to the proof of (\ref{eqB.16}), we can prove that 
\begin{eqnarray}
&&\max_{b_k-\varepsilon\varpi_k\leq t< b_k}\frac{1}{|{\cal C}(b_k)|^{1/2}}\left\vert\left[D_k(t)-D_k(b_k)\right]\sum_{i\in{\cal C}(b_k)}\int_{\mathbb C}Z_{iT}^\epsilon(b_k/T;u)\delta_i(u)du\right\vert\notag\\
&=&O_P\left(\frac{\varpi_k}{(T|{\cal C}(b_k)|)^{1/2}}\left(\sum_{i\in{\cal C}(b_k)}\Vert\delta_i\Vert^2\right)^{1/2}\right)\notag\\
&=&O_P\left(\frac{1}{|{\cal C}(b_k)|^{1/2}}\left(\frac{1}{|{\cal C}(b_k)|^{1/2}}\sum_{i\in{\cal C}(b_k)}\Vert\delta_i\Vert^2\right)^{1/2}\right).\label{eqB.23}
\end{eqnarray}
On the other hand, observe that
\[Z_{iT}^\epsilon(t/T,u)-Z_{iT}^\epsilon(b_k/T,u)=\frac{1}{\sqrt{T}} \left[-\sum_{s=t+1}^{b_k} \epsilon_{is}(u)+\frac{b_k-t}{T}\sum_{s=1}^T\epsilon_{is}(u)\right]\]
for $b_k-\varepsilon\varpi_k\leq t< b_k$, and 
\begin{eqnarray}
&&\max_{b_k-\varepsilon\varpi_k\leq t< b_k}\left\vert \frac{D_k(t)}{(T|{\cal C}(b_k)|)^{1/2}}\sum_{i\in{\cal C}(b_k)} \sum_{s=t+1}^{b_k}\int_{\mathbb C}\epsilon_{is}(u)\delta_i(u)du\right\vert\notag\\
&=&O_P\left(\frac{\varpi^{1/2}_k}{|{\cal C}(b_k)|^{1/2}}\left(\sum_{i\in{\cal C}(b_k)}\Vert\delta_i\Vert^2\right)^{1/2}\right)\notag\\
&=&O_P\left(\frac{T^{1/4}}{|{\cal C}(b_k)|^{3/8}}\left(\frac{1}{|{\cal C}(b_k)|^{1/2}}\sum_{i\in{\cal C}(b_k)}\Vert\delta_i\Vert^2\right)^{1/2}\right),\label{eqB.24}
\end{eqnarray}
and
\begin{eqnarray}
&&\max_{b_k-\varepsilon\varpi_k\leq t< b_k}\left\vert \frac{(t-b_k)D_k(t)}{|{\cal C}(b_k)|^{1/2}T^{3/2}}\sum_{i\in{\cal C}(b_k)}^N \sum_{s=1}^{T}\int_{\mathbb C}\epsilon_{is}(u)\delta_i(u)du\right\vert\notag\\
&=&O_P\left(\frac{\varpi_k}{(|{\cal C}(b_k)|T)^{1/2}}\left(\sum_{i\in{\cal C}(b_k)}\Vert\delta_i\Vert^2\right)^{1/2}\right)\notag\\
&=&O_P\left(\frac{1}{|{\cal C}(b_k)|^{1/2}}\left(\frac{1}{|{\cal C}(b_k)|^{1/2}}\sum_{i\in{\cal C}(b_k)}\Vert\delta_i\Vert^2\right)^{1/2}\right).\label{eqB.25}
\end{eqnarray}
By (\ref{eqB.23})--(\ref{eqB.25}) and noting that $T=O(|{\cal C}_\bullet|^{3/2})$, we can show that the fourth term on RHS of (\ref{eqB.21}) is $o_P\left(\frac{1}{|{\cal C}(b_k)|^{1/2}}\sum_{i\in{\cal C}(b_k)}\Vert\delta_i\Vert^2\right)$ uniformly over $b_k-\varepsilon\varpi_k\leq t< b_k$. By (\ref{eqB.14}), with probability approaching one, the first two terms on RHS of (\ref{eqB.21}) are bounded uniformly over $b_k-\varepsilon\varpi_k\leq t< b_k$. By (\ref{eqB.19}) and (\ref{eqB.20}), we may show that the third term on RHS of (\ref{eqB.21}) is $o\left(\frac{b_k-t}{|{\cal C}(b_k)|^{1/2}}\sum_{i\in{\cal C}(b_k)}\Vert\delta_i\Vert^2\right)$ for any $b_k-\varepsilon\varpi_k\leq t< b_k$. Combining the above arguments, we prove that the fourth term on RHS of (\ref{eqB.21}) is the asymptotically dominated one. Then, we may conclude that
\begin{equation}\label{eqB.26}
{\sf P}\left(\max_{b_k-\varepsilon\varpi_k\leq t< b_k}Z(t; {\cal C}(b_k))< Z(b_k; {\cal C}(b_k))\right)\rightarrow1.
\end{equation} 
In exactly the same way, we can also show that 
\begin{equation}\label{eqB.27}
{\sf P}\left(\max_{b_k< t\leq b_k+\varepsilon\varpi_k}Z(t; {\cal C}(b_k))< Z(b_k; {\cal C}(b_k))\right)\rightarrow1.
\end{equation} 
With (\ref{eqB.26}) and (\ref{eqB.27}), we complete the proof of Proposition A.5.\hfill$\Box$

\end{document}